\documentstyle[epsfig]{aipproc}
\begin{document}
\def\be{\begin{equation}}
\def\ee{\end{equation}}
                         \def\bearr{\begin{eqnarray}}
                         \def\eearr{\end{eqnarray}}
\def\benum{\begin{enumerate}}
\def\eenum{\end{enumerate}}
\def\bitem{\begin{itemize}}
\def\eitem{\end{itemize}}
\def\bedes{\begin{description}}
\def\edes{\end{description}}
                         \def\eg{ {\em e.g.}~}
                         \def\etal{ {\em et al.}~}
                         \def\ie{ {\em i.e.}}
                         \def\viz{ {\em viz.}~}
\def\lsim{\:\raisebox{-0.5ex}{$\stackrel{\textstyle<}{\sim}$}\:}
\def\gsim{\:\raisebox{-0.5ex}{$\stackrel{\textstyle>}{\sim}$}\:}
\def\go{\rightarrow}
\def\goes{\longrightarrow}
\def\hrar{\hookrightarrow}
\def\bul{\bullet}
\def\mET{E_T \hspace{-1.1em}/\;\:}
\def\mpT{p_T \hspace{-1em}/\;\:}
\def\rpv{$R_p \hspace{-1em}/\;\:$}
\def\bv{$B \hspace{-1em}/\;\:$}
\def\rp{$R$-parity}
\def\tb{\tan \beta}
                         \def\N0{\widetilde \chi^0}
                         \def\Chip{\widetilde \chi^+}
                         \def\Chim{\widetilde \chi^-}
                         \def\Chipm{\widetilde \chi^\pm}
                         \def\Chimp{\widetilde \chi^\mp}
\def\l{\lambda}
\def\lp{\lambda'}
\def\lpp{\lambda''}
                         \def\sq{\widetilde q}
                         \def\su{\widetilde u}
                         \def\sd{\widetilde d}
                         \def\sc{\widetilde c}
                         \def\ss{\widetilde s}
                         \def\st{\widetilde t}
                         \def\sb{\widetilde b}
\def\sl{\widetilde \ell}
\def\sel{\widetilde e}
\def\snu{\widetilde \nu}
\def\smu{\widetilde \mu}
\def\stau{\widetilde \tau}
\def\glu{\widetilde g}
\def\tm{\widetilde m}
\def\epem{e^+ e^-}
\def\emem{e^-e^-}

\title{SUSY and  SUSY Breaking Scale at the Linear Collider}

\author{R. M. Godbole}
\address{Centre for Theoretical Studies, \\ Indian Institute of Science,\\
Bangalore, 560 012 \\ INDIA }

\begin{flushright}
IISc-CTS/02/01 \\
hep-ph/0102191 
\end{flushright}

\vskip 25pt
\begin{center} 
{\large \bf SUSY and  SUSY Breaking Scale at the Linear Collider
\footnote{Plenary talk presented  at LCWS 2000, Fermilab, 
Oct. 26-30, 2000}}    
       \\
\vskip 25pt

{\bf                        Rohini M. Godbole } \\ 

{\footnotesize\rm 
                      Centre for Theoretical Studies, \\
                     Indian Institute of Science, \\
                      Bangalore 560 012, India. \\ 
                     E-mail: rohini@cts.iisc.ernet.in  } \\ 

\vskip 20pt

{\bf                             Abstract 
}

\end{center}

\begin{quotation}
\noindent
After summarising very briefly the key features of different model
predictions for sparticle masses and their relation with the supersymmetry
(SUSY) breaking scales and parameters, I discuss the capabilities of an $e^+
e^-$ Linear Collider (LC) with $\sqrt{s} \geq$ 500 GeV for precision
measurements of sparticle properties. Then I focus on the lessons one can
learn about the scale and mechanism of SUSY breaking from these measurements
and point out how LC can crucially complement and extend the achievements of
the LHC. I end by mentioning what would be the desired extensions in the
type/energy of the colliding particles and their luminosity from the point
of view of SUSY investigations. 
\end{quotation}

\newpage

\maketitle

\begin{abstract}
After summarising very briefly the key features of different model
predictions for sparticle masses and their relation with the supersymmetry
(SUSY) breaking scales and parameters, I discuss the capabilities of an $e^+
e^-$ Linear Collider (LC) with $\sqrt{s} \geq$ 500 GeV for precision
measurements of sparticle properties. Then I focus on the lessons one can
learn about the scale and mechanism of SUSY breaking from these measurements
and point out how LC can crucially complement and extend the achievements of
the LHC. I end by mentioning what would be the desired extensions in the
type/energy of the colliding particles and their luminosity from the point
of view of SUSY investigations. 

\end{abstract}

\section*{Introduction}
In this talk I essentially want to discuss how a linear collider (LC) will
do the dual job of aiding to establish supersymmetry (SUSY) as a viable
theory and giving information about the scale of SUSY breaking  along with
pointing way towards an understanding of the mechanism of SUSY breaking.

The testing of the Standard Model (SM) to an unprecedented accuracy has
confirmed the correctness of the SM as a renormalisable gauge field theory,
at least as an effective theory. This has increased the attraction of TeV
scale SUSY even further. It is the only concrete and completely worked out
mechanism we have which stabilizes the Higgs boson mass $m_h$ at the electroweak
scale \footnote { Of course, `warped large' extra dimensions~\cite{RS1} 
might obviate the heirarchy problem completely.} and provides a natural 
mechanism for the spontaneous breakdown of the EW symmetry.

However, in spite of all these theoretical attractions of the SUSY, the only
indication of its possible existence we have is the (non)unification of the
SU(3), SU(2) and U(1) gauge couplings, when evolved from their accurately
measured low energy values, at a very high scale in the MSSM(SM), as shown 
\begin{figure}[htb]
\centerline{
 \includegraphics*[scale=0.35]{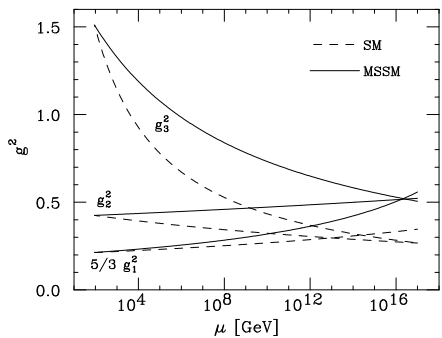}
}
\caption{(Non)Unification of the three couplings in the (SM) MSSM.}
\label{F:rgplen:0}
\end{figure} 
here in Fig.~\ref{F:rgplen:0}. Our
theoretical understanding of the SUSY breaking mechanism, though enriched in
recent years, is not `really' complete~\cite{HM2P}. In almost all of the 
formulations, SUSY is broken dynamically at a high scale and then this 
breaking is mediated to our low energy world. Various Soft Supersymmetry 
Breaking (SSB) parameters at the high scale of SUSY breaking are decided by 
the choice of the SSB mechanism and the mediation mechanism. Various 
theoretical and
experimental considerations restrict the scale to a rather big range $10^4
GeV$ $<$ $M_{SSB}$ $<$ $M_{Pl}$. The low energy values of the SUSY breaking
parameters are then decided by the renormalization group evolution. Thus the
sparticle masses and in cases where mixing occurs even their couplings,
depend on the SSB mechanism. These low energy values of sparticle properties
(if we find them) are the only clues available to us to point towards the
physics at high scale and hence at the SUSY breaking mechanism; just like
the measurement of low energy gauge couplings have offered a possible
telescope at the unification~(cf. Fig.~\ref{F:rgplen:0}).

In the early days of SUSY models there existed essentially only one 
one class of models where the SSB is transmitted via gravity to the
low energy world. In the past few years
there has been tremendous progress in the ideas about SUSY breaking and thus 
there exist now a set of different models:
\bedes
\item[1)] Gravity mediated models include models like (minimal) SUGRA
(mSUGRA), (constrained) MSSM (cMSSM) etc.  The difference between mSUGRA 
and cMSSM in my notation is the fact that $\mu$ is determined by the 
condition for the radiatively induced spontaneous symmetry breakdown of
EW symmetry to occur in the former, whereas in the latter it is a 
free parameter. Both assume universality of the gaugino and sfermion 
masses  at the high scale.  In this  case supergravity couplings 
of the fields in the hidden sector with the SM fields are responsible for 
the soft SSB terms.  These models always have  extra scalar mass parameter 
$m_0^2$ which  needs fine tuning so that the sparticle exchange does not 
generate FCNC effects, at an unacceptable level.  
\item[2)]In the Anomaly Mediated Supersymmetry Breaking (AMSB) models 
supergravity couplings which cause mediation are absent and the 
SSB is caused by loop effects. The conformal anomaly generates the 
soft SSB terms in this case and the sparticles acquire masses due to the
breaking of scale invariance. Note that this contribution exists 
even in the case of mSUGRA/MSSM, but is much smaller
in comparison with the tree level terms which exist in those models. 
This mechanism becomes a viable one  for solely generating the SSB 
terms, when the quantum contributions to the gaugino masses
due to the `superconformal anomaly' can be large~\cite{RS2,GR}, hence
the name Anomaly mediation for them.  The slepton masses in this model are 
tachyonic in the absence of a scalar mass parameter $m_0^2$.
\item[3)] An alternative scenario where the SSB is transmitted to the low
energy world via a  messenger sector   through  messenger fields which have 
gauge interactions, is called the Gauge Mediated Supersymmetry Breaking 
(GMSB)~\cite{gmsb_rev}. These models have no problems with the FCNC 
and do not involve any scalar mass parameter.
\item[4)]  There exist also a  class of models where the mediation of the
symmetry breaking is dominated by gauginos~\cite{gaumsb}. In these
models the wave function of the matter particles and their superpartners
at the SUSY breaking brane is suppressed, whereas  those
of the gauginos is substantial, due to the fact that the gauge superfields
live in the bulk. Hence the matter sector feels the effects of SUSY breaking
dominantly via gauge superfields. As a result, in these scenarios, one expects 
$m_0 \ll m_{1/2}$, reminiscent of the `no scale' models.

\edes
All these models clearly differ in their specific predictions for various 
sparticle spectra, features of some of which are summarised in 
\begin{table}[htb]
\caption{The table gives predictions of different types of SUSY breaking
models for gravitino, gaugino and scalar masses $\alpha_{i} =
{g_{i}^{2}}/{4 \pi}$ (i=1,2,3 corresponds to U(1), SU(2) and SU(3)
respectively), $b_{i}$ are the coefficients of the
${-g_{i}^{2}}/{(4 \pi )^{2}}$ in the expansion of the $\beta$ functions 
$\beta_{i}$ for  the coupling $g_i$ and $a_i$ are the coeffecients 
of the corresponding expansion of the anomalous dimension. the coeffecients 
$D_i$ are the squared gauge charges multiplied by various factors which 
depend on the loop contributions to the scalar masses in the different models.}
\vspace{0.2cm}
\begin{tabular}{cccc}
 Model & $m_{\tilde{G}}$ & $(mass)^2$ for gauginos & $(mass)^2$ for scalars\\
&&&\\
\tableline
&&&\\
 mSUGRA & ${{M_{SSB}^{2}} / {\sqrt{3} M_{pl}}}$ $\sim $ TeV &
$({\alpha_{i}}/{\alpha_{2}})^{2}$ $M_{2}^{2}$ & $m_{0}^{2} + \sum_{i}
D_{i} M_{i}^{2}$ \\
cMSSM & $M_{SSB} \sim 10^{10} - 10^{11}$ GeV & $\mbox{ }$ & $\mbox{  }$ 
\\
&&&\\
\hline
&&&\\
GMSB & $({\sqrt{F}}/{100TeV})^{2}$ eV &
$({\alpha_{i}}/{\alpha_{2}})^{2} M_2^2$ & $\sum_{i} D_{i}^{'} M_{2}^{2}$ \\
$\mbox { }$  & 10 $<  \sqrt{F} < 10^4 $ TeV &  & \\ 
&&&\\
\hline
&&&\\
AMSB & $\sim $ 100 TeV & 
$({\alpha_{i}}/{\alpha_{2}})^{2} ({b_{i}}/{b_{2}})^{2} M_2^2$ & 
$\sum_{i} 2 a_{i} b{i} ({\alpha_{i}}/{\alpha_{2}})^{2} M_2^2$ \\ 
&&&\\
\hline
\end{tabular}
\label{T:rgplen:1}
\end{table}
Table~\ref{T:rgplen:1} following ~\cite{peskin_talk}, where the usual
messenger scale parameter $\Lambda$ had been traded for $M_2$ for ease of
comparison. As one can see the 
expected mass 
of the gravitino varies widely in different models. The SUSY breaking scale
$\sqrt{F}$ in GMSB model is restricted to the range shown in Table 1 by
cosmological considerations. Since $SU(2), U(1)$ gauge groups are not
asymptotically free,\ie, $b_i$ are negative, the slepton masses are tachyonic
in the AMSB model, without a scalar mass parameter, as can be seen from 
the third column of the 
table. The minimal cure to this is, as mentioned before, to add an additional 
parameter $m_{0}^{2}$, not shown in the table, which however spoils
the RG invariance.  In  the 
gravity mediated models like mSUGRA, cMSSM and most of the versions of 
GMSB models, there exists gaugino mass unification at high scale, whereas 
in the AMSB models the gaugino masses are given by RG invariant equations 
and hence are determined completely by the values of the couplings at 
low energies and become ultraviolet insensitive. Due to this very different 
scale dependence,
the ratio of gaugino mass parameters at the weak scale in the two sets of
models are quite different: models I and II have $M_1 : M_2 : M_3$ = 1 : 2 :
7 whereas in the AMSB model (III) one has $M_1 : M_2 : M_3$ = 2.8 : 1 : 8.3.
The latter therefore, has the striking prediction that the lightest chargino
$\Chipm_{1}$ and the LSP $\N0_{1}$, are almost  pure
SU(2) gauginos and are very close in mass. The expected particle spectra in
any given model can vary a lot. But still one can make certain general
statements,\eg the ratio of squark masses to slepton masses is usually 
larger in the GMSB models as compared to mSUGRA.
In mSUGRA one expects the sleptons to be lighter than the first two
generation squarks, the LSP is expected mostly to be a bino and the right
handed sleptons are lighter than the left handed sleptons. On the other
hand, in the AMSB models, the left and right handed sleptons are almost
degenerate. The above mentioned degeneracy between $\tilde{\chi}_{1}^{\pm}$
and $\tilde{\chi}_{1}^{0}$ is lifted by the loop effects~\cite{extra1}. For
$\Delta M$ = $m_{\tilde{\chi}_{1}^{\pm}}$ - $m_{\tilde{\chi}_{1}^{0}}$
$<$ 1 GeV, the phenomenology of the sparticle searches in AMSB models will be
strikingly different from that in mSUGRA, MSSM etc. In the GMSB models, 
the LSP is gravitino and is indeed `light' for the range of the values of
$\sqrt{F}$ shown in Table~\ref{T:rgplen:1}. The candidate for the next lightest
sparticle, the NLSP can be $\N0_{1}$, $\stau_{1}$ or
$\sel_R$ depending on model parameters. The NLSP life times
and hence the decay length of the NLSP in lab  is given by 
$ L = c\tau \beta \gamma \propto \frac{1}{(M_{LSP})^5}$ $(\sqrt{F})^{4}$. 
Since the theoretically 
allowed 
values of $\sqrt{F}$ span a very wide range as shown in Table 1, so do those
for the expected life time and this range is given by 
$10^{-4}$ $<$ c$\tau \gamma \beta$ $<$ $10^{5}$ cm.
Since the crucial differences in different models exist in the slepton
and the chargino/neutralino sector,  it is clear that the leptonic 
colliders which can study these sparticles with the EW interactions, with
great precision, can  play really a crucial role in being able to distinguish
among different models.

The above discussion, which illustrates the wide `range' of predictions of
the SUSY models,  also makes it clear that  a general discussion of the  
sparticle phenonenology  at any collider is far too complicated. To me, 
that essentially reflects our ignorance. This makes it even more imperative 
that we try to extract as much model independent information from the 
experimental measurements. This is one aspect where the leptonic 
colliders can really play an extremely important role.

\vspace{0.3cm}
\noindent\underline{{\it Questions about SUSY we need answered by next 
generation colliders}}

\noindent 
We need the next generation colliders to first establish SUSY as a viable
theory and further extract information about the SUSY breaking mechanism and
scale. In particular, we need to

1) {\bf Find the sparticles} and establish their quantum numbers.

2) The latter can be done  only by checking the interactions of sparticles
and establish coupling equalities implied by the symmetry.

3) Determine the {\bf scalar masses, gaugino masses} and gaugino-higgsino
mixing.

4) Measure the properties of the third generation sfermions including the
L-R mixing.

The measurements mentioned in (3) above can give information about $\mu$,
tan$\beta$ and some of the soft SUSY breaking parameters whereas (4) above can
further add to the determination of $\mu$, tan$\beta$, trilinear A
parameters and the scalar mass parameters. The LHC will be able to achieve
the goals given in `bold face' in the list above; for the remaining tasks we
need the clean environment of the $e^+e^-$ colliders.

\vspace{0.5cm}
\noindent\underline{{\it What LHC can do}}

\noindent 
Let us start with a  summary of major 
hopes~\cite{cms_rep,atlas_tdr,lhc_susy,extra2} from LHC for SUSY enthusiasts.
Various versions of `naturalness' arguments~\cite{barb,anderson,moroi} 
indicate that if theories are `natural', at least some of the sparticles, 
notably the gauginos/higgsinos, must be accessible at LHC. Thus if SUSY 
is realized in  nature, LHC should be able to provide some proof for it. 
Being a hadronic collider, LHC is best suited
for the search of strongly interacting particle sector. The heavier `strongly
interacting' sparticles will be produced first and the lighter sparticles
with EW interactions only in the decay. The very high rates~\cite{tata1}
(\eg, even for a gluino mass  of 2 TeV, the expected cross-section 
is $\sim 10$ fb, giving about 1000 events for the high luminosity option)
make discovery easy. Methods have been developed to make accurate measurements
of different sparticle masses; a nontrivial task as the worst background for 
SUSY searches is SUSY itself~\cite{paige1}. 
Depending on the point in mSUGRA parameter space chosen for analysis, a 
determination of $m_{\sq_L}, m_{\glu}$ upto an accuracy of $5-7 \% $  
is possible; whereas the  masses $m_{\N0_1}, m_{\N0_2}$ can be determined 
with $ < 10\%$ accuracy~\cite{atlas_tdr,lhc_susy,tata1,paige1}. For some of the 
points chosen for studies high accuracies $\sim 1-2 \%$ are also possible
for neutralino mass determination. 
Ingenious methods have been eveloped to get an idea of the effective 
SUSY breaking scale~\cite{paige1}. However, accurate information about
the SUSY breaking scale {\bf and} mechanism generally does not seem easily
extractable. Further, a direct determination of quantum numbers and couplings
of the sparticles is not possible. The heavier gauginos are not accessible as
the rates for direct, EW production are very low. The reach for sleptons at
LHC is limited as compared to that for the strongly interacting 
particles and is  $m_{\tilde{l}}$ $\leq$ 360 GeV unless it is produced
in cascades of squarks; a model dependent fact. It has been shown that many
SUSY model parameters such as $\mu$, tan$\beta$, $M_2$, $M_3$ can be
determined with an accuracy of a percent level~\cite{paige1,atlas_tdr1},  
within a model. However,
model independent analyses do not yet promise similar accuracy~\cite{paige2}. 
Further, if we want the LHC measurements to provide us with a clue about 
the nature of the dark matter in the universe, it will be possible  
only if $m_{\tilde{l}_{R}}$ $<$ $m_{\tilde{\chi}_{2}^{0}}$. These analyses
essentially need determination of the chargino/higgsino content of
$\tilde{\chi}_{i}^{0}$. At LHC this is possible only if $m_{\tilde{l}_{R}}$
$<$ $m_{\tilde{\chi}_{2}^{0}}$, as has been recently 
demonstrated~\cite{man-mih}. This is one area where a leptonic collider 
can make very crucial contributions. As a matter of fact, this information, 
if available, can play a very useful role in LHC analyses too. Thus 
information obtained from an LC can feed back into LHC analyses.

\vspace{0.5cm}
\noindent\underline{{\it What do we expect an LC with $500 < \sqrt{s} < 1000$
GeV to tell us about SUSY?}}

\begin{figure}[ht]
\centerline{
 \includegraphics*[width=2.8in,height=2.5in]{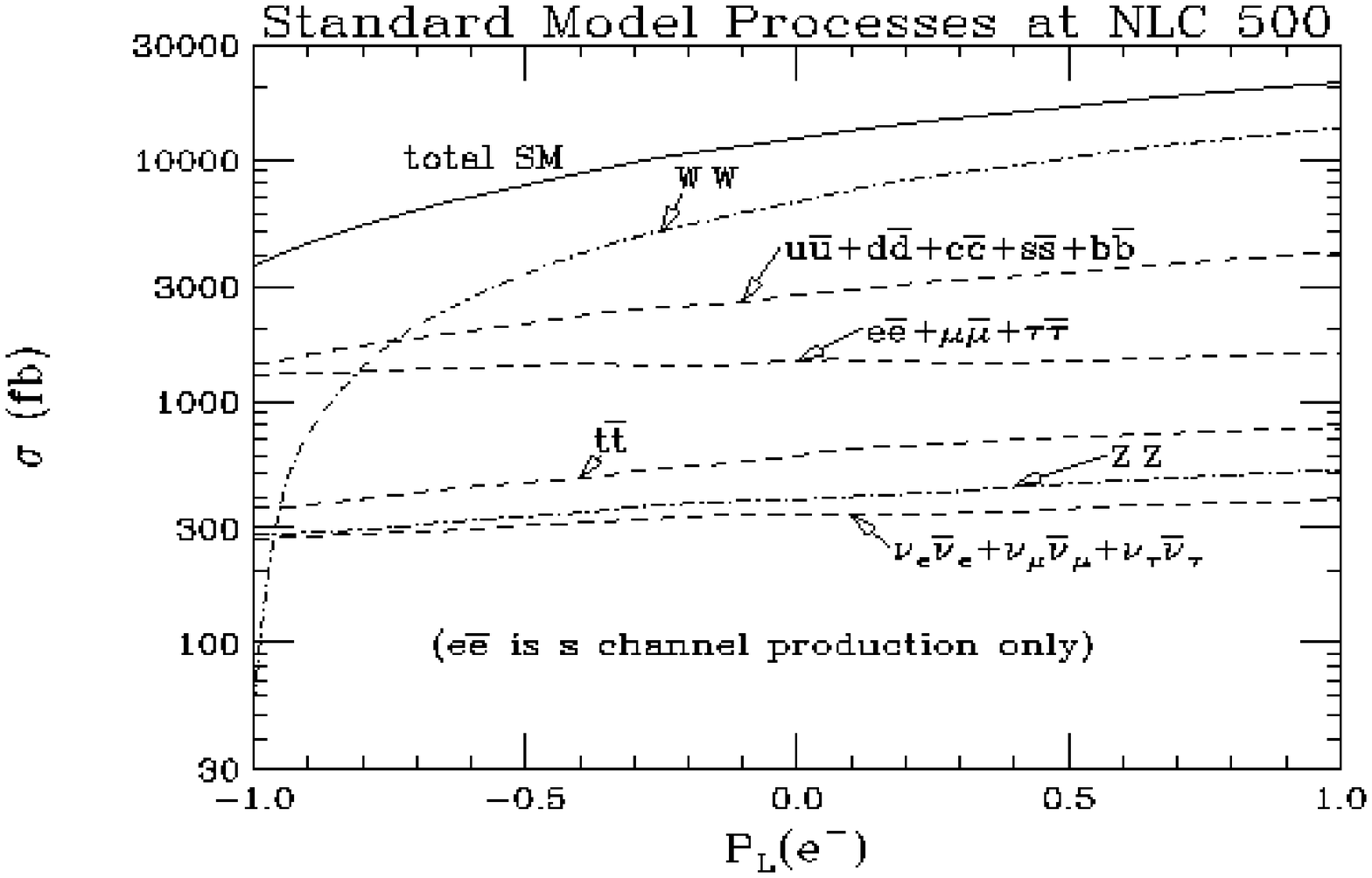}
 \includegraphics*[width=2.8in,height=2.5in]{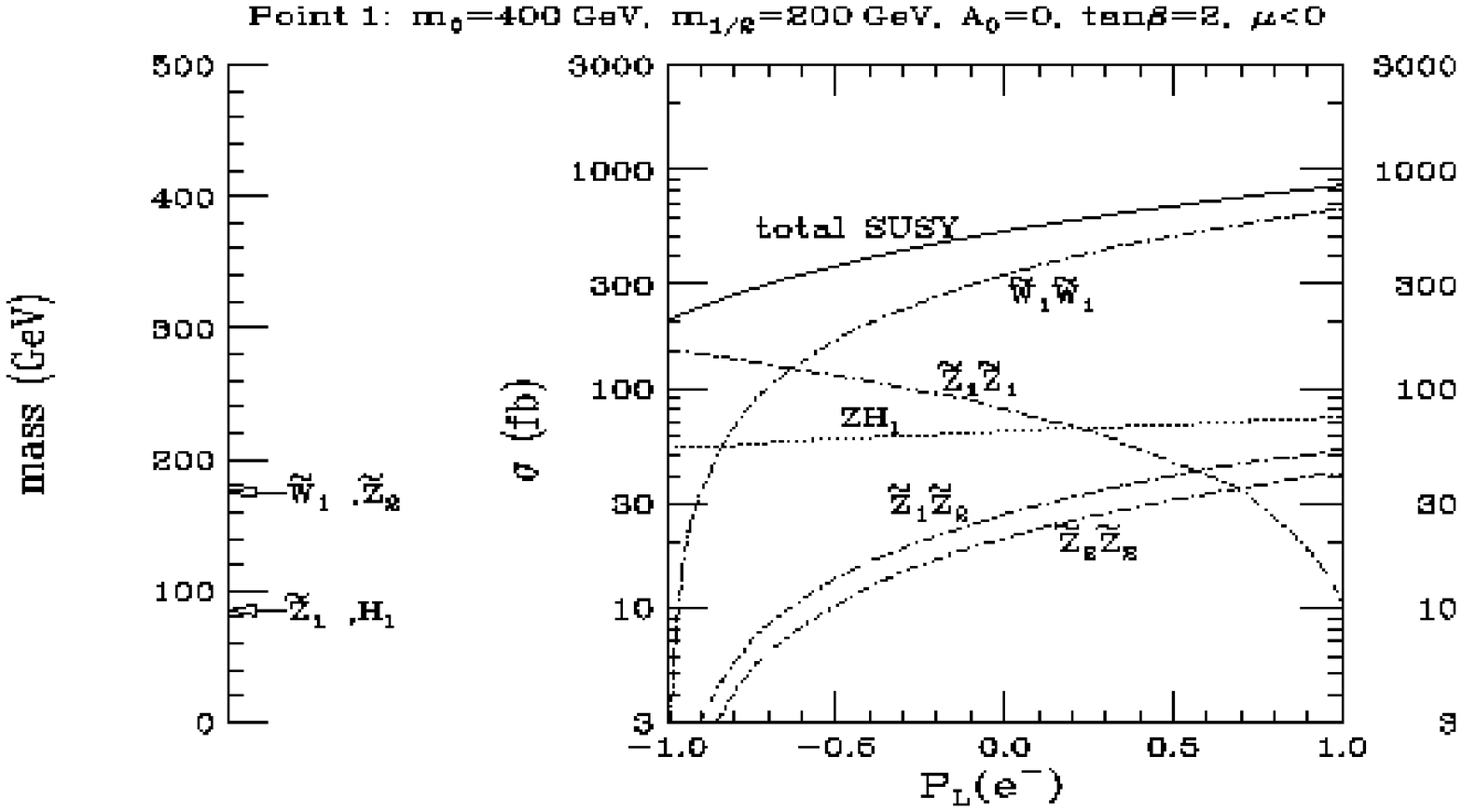}
}
\caption{Cross sections of different SM processes as well as the 
chargino/neutralino production. The values of the model parameters
are mentioned in the figure.}
\label{F:rgplen:1}
\end{figure}

\noindent The above discussion identifies the expectations from LC from 
the point of view of SUSY as follows: 
\bedes
\item[1)] An LC should provide {\bf precision} measurement of sparticle 
masses and mixing. Of course for that one needs $\sqrt{s}$ $>$ 2$m_{s}$,
where $m_{S}$ stands for sparticle mass and thus the desirable energy  range
for an LC from the point of view of SUSY searches should extend at least 
upto 1000  GeV.
\item[2)] An LC should provide determination of quantum numbers such as 
spin, hypercharge and establish the equality of couplings predicted by SUSY.
\item[3)] Information from LHC, alongwith measurements in (2) can then be 
used to get information about the SUSY breaking at high scale.
\edes
As seen before, LHC can achieve the first goal only partially and the second
one only indirectly. The information on sparticle masses obtained from LHC
can serve as an important input to choose energies at which to run the LC.
The tunable energy of an $e^+e^-$ LC allows for sequential production of
various sparticles and hence a better knowledge of the possible SUSY
background to SUSY search.  Since SUSY involves chiral fermions and their
spartners, polarisation of the initial $e^+$/$e^-$ beam can be used  very
effectively to project out information about particle spectra and couplings.
Appropriate choice of polarisation can also reduce effectively the 
background due to $W^+W^-$  production which has a very high rate.
Fig.~\ref{F:rgplen:1} taken from Ref.~\cite{tata2} shows the
cross-sections for different SM processes and the corresponding ones for the
SUSY model dependent chargino/neutralino pair production, at a chosen  point 
in the mSUGRA parameter space. From the figure it is clear that with a judicious
choice of polarization of $e^-$/$e^+$ beam, the SM background can be handled
and precision measurements of chargino/neutralino sector are possible. The
$e^+e^-$ collider produces democratically all the sparticles that have EW 
couplings. Hence it is better suited than the LHC to study the 
gauginos/higgsinos and sleptons and will complement the information in
these sectors from the LHC very effectively. The correlation between 
properties of gluino that will be obtained from the LHC and those of
chargino/neutralino sector from LHC/LC can disentangle the various gaugino
mass parameters $M_i$ at weak scale. For reasons outlined in the introduction 
knowledge about the relative values of $M_i (i=1,3)$ at the weak scale, 
from independent sources, contains crucial clues to the physics at high
scale. This also shows how truly the LHC and LC are complementary to each
other and thus how necessary both are to solve the puzzle of EWSB.

\section*{What and How well can LC measure}
There have been a large number of dedicated 
studies~\cite{tata2,MP1,extra3,ACC,extra4,extra5,teslatdr,zerwastlk,extra6} 
of the possibilities of
precision measurements of the 
sleptons~\cite{extra6,FM1,MIHO1,tata2,MIHO2,MB1,SM}, 
squarks~\cite{FF,tata2,SM,DEKG,stop_mc}, 
charginos/neutralinos~\cite{extra6,tata2,FM1,FS,MB1,Z1,Z2,Z3,extra7,MGP1,kneur1,kneur} and Higgses~\cite{MB}. Study of third generation 
sfermions~\cite{MIHO1,MIHO2,SM} are shown to yield particularly 
interesting information about SUSY models. Almost no study is possible 
without use of polarisation, at least for one of the  initial state fermions.
A detailed discussion of the special advantages, in general, of using 
polarisation of  {\it both} the beams is available elsewhere in the
proceedings~\cite{gtalk}.
In this talk I do not include discussion of the SUSY Higgses as it is also
discussed somewhere else in the proceedings~\cite{MB} and also 
because the dependence of the Higgs sector on SUSY breaking parameters 
and hence on the high scale physics is essentially only through  the loop 
corrections. 
\subsection*{Precision measurements of masses}
\noindent\underline{{\it Sleptons and Charginos/Neutralinos}}

\begin{figure}[ht]
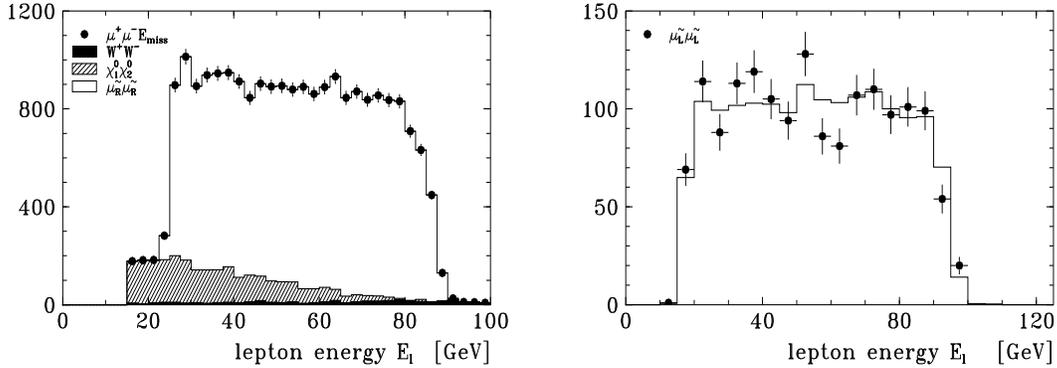

\centerline{
 \includegraphics*[scale=0.35]{fig2l.eps}
 \includegraphics*[scale=0.35]{fig2r.eps}
}
\caption{Precision of mass determination for $\smu_{L}$ and $\smu_{R}$ at TESLA
with $500~{\rm fb}^{-1}$,  with $P_{e^-}/P_{e^+} = 0.8/0.6$, taken from the 
studies in \protect\cite{MB1}. 
Details of the values of the mSUGRA parameters are given there.}
\label{F:rgplen:2}
\end{figure}

\noindent The masses of sleptons can be determined at an LC essentially using
kinematics. Making use of partial information from the LHC, 
it will be
possible to tune the energy of the LC to produce the sfermions sequentially.
The pair produced lightest sleptons will decay through a two body decay.
Let us take the example of $\smu_R$ which will have the simplest decay.
So one has in this case, 
\be
e^+e^- \to \tilde{\mu}_{R} \tilde{\mu}_{R}^{*} \to \mu^+ \mu^-
\tilde{\chi}_{1}^{0} \tilde{\chi}_{1}^{0}
\ee
Since the slepton is a scalar the decay  energy distribution for the
$\mu$ produced in two body decay of $\smu_R$, will be flat with 
\be
{m_{\smu_R} \over 2} \left(1-{{m_{\N0_1}^2} \over {m_{\smu_R}^2}} \right)
 \gamma (1-\beta) < E_\mu <
{m_{\smu_R} \over 2} \left(1-{{m_{\N0_1}^2} \over {m_{\smu_R}^2}}\right)
\gamma (1+\beta) .
\ee
Thus measuring the end points of the $E_{\mu}$ spectrum accurately will
yield a precision measurement of the masses $m_{\smu_{R}}$,
$m_{\N0_1}$. Of course, one has to contend with the background from $W^+W^-$ 
and  the $\N0_1 \N0_2$ production (cf. Fig.~\ref{F:rgplen:1}). 
As can be seen from the right panel of the same figure,
this can be handled by choosing polarised $e^-/e^+$ 
beams.  Fig.~\ref{F:rgplen:2} taken from Ref.~\cite{MB1} shows that
$\tilde{\mu}_{R}$ mass can be determined to a precision of 0.3 \% at TESLA
with $\int {\cal L} dt$ = 500 f$b^{-1}$. The analysis uses {\bf both}
$e^-$/$e^+$ beam polarisations, with $P_{e^-}/P_{e^+} = 80\%/60\%$.
The need for polarisation of {\it both} 
the beams  is discussed elsewhere in the proceedings~\cite{gtalk}.
The panel on right shows, for the same point in mSUGRA parameter space as
the panel in left, signal for production of $\smu_{L}^{*} \smu_{L}$ and 
its three body decay via 
\be
e^+e^- \to \smu_{L} \smu_{L}^{*} \to \mu^{-} \N0_{2} \mu^{+} \N0_{2} 
\to \mu^{-} \mu^{+} \N0_{1} \N0_{1} l^{+} l^{-} {l'}^{+} {l'}^{-}
\ee
For the particular point in the mSUGRA parameter space they have chosen,
the branching ratio  for  $\smu_{L} \to \mu \N0_{2}$ is substantial
and the cleanliness of the final state compensates  for the eventual small
rates.  Thus $\smu_{L}$ mass determination to percent level seems possible  
for TESLA~\cite{MB1}.  However, the method of using the end 
point of the energy spectrum will not work so well,\eg ,  
for $\stau_{1} \stau_{1}^{*}$ production and decay. 

Another method for precision determination of the masses of the sleptons and 
the lighter charginos/neutralinos, is to perform threshold scan. The 
linear $\beta$
dependence as opposed to the $\beta^{3}$ dependence of the cross-section,
near the threshold (where $\beta$ is the c.m. velocity of the produced
sparticle) makes the method more effective for the spin 1/2
charginos/neutralinos than the sleptons. Of course, such threshold scans
will require very high luminosity. The efficacy of the method of threshold
scans has been studied in the context of the high luminosity TESLA collider
~\cite{MB1}. The results of their study for a chosen point in the 
mSUGRA parameter space are summarized in Table ~\ref{T:rgplen:2}. 
\begin{table}
\caption{Kinematic mass determinations at TESLA for threshold scans with
$10$ fb $^{-1}$ luminosity per energy, for the mSUGRA point RR1 of the TESLA 
studies~\protect\cite{teslatdr}.}
\label{T:rgplen:2}
\begin{tabular}{ccccc}
particle & m & $\delta m_{cont}$ &$\delta m_{scan}$ & Can give info.on \\ 
\tableline
  $\widetilde \mu_{R}$ & 132.0 & 0.3 & 0.09 & $m_{0}$, $m_{1/2}$, $\tb$ \\
  $\widetilde \mu_{L}$ & 176.0 & 0.3 & 0.4 & $\mbox{ }$ \\
  $\widetilde \nu_{\mu}$ & 160.6 & 0.2 & 0.8 & $\mbox{ }$ \\
  $\widetilde e_{R}$ & 132.0 & 0.2 & 0.05 & $\mbox{ }$ \\
  $\widetilde e_{L}$ & 176.0 & 0.2 & 0.18 & $\mbox{ }$ \\
  $\widetilde \nu_{e}$ & 160.6 & 0.1 & 0.07 & $\mbox{ }$ \\ \hline
  $\widetilde \tau_{1}$ & 131.0 & $\mbox{ }$ & 0.6 & $m_{0}$, $m_{1/2}$,$\mu$, $\tb$ \\
  $\widetilde \nu_{\tau}$ & 160.6 & $\mbox{ }$ & 0.6 & $\mbox{ }$ \\ 
\tableline
  $\Chipm_{1}$ & 127.7 & 0.2 & 0.04 & $M_{2}$, $\mu$, $\tb$ \\
  $\Chipm_{2}$ & 345.8 & $\mbox{ }$ & 0.25 & $\mbox{ }$ \\
\tableline
  $\N0_{1}$ & 71.9 & 0.1 & 0.05 & $M_{1}$,$M_{2}$, $\mu$, $\tb$ \\
  $\N0_{2}$ & 130.3 & 0.3 & 0.07 & $\mbox{ }$ \\
  $\N0_{3}$ & 319.8 & $\mbox{ }$ & 0.30 &  $\mbox{ }$ \\
  $\N0_{4}$ & 348.2 & $\mbox{ }$ & 0.52 & $\mbox{ }$ \\
\tableline
\end{tabular}
\end{table}
This requires about $100~{\rm fb}^{-1}$ luminosity distributed over 
10 energy values for each sparticle. For the $\snu_e$ the high accuracy 
of the mass determination is possible because of the large cross-section
due to the $t$ channel contribution. For the $\snu_{\mu}$ and $\snu_{\tau}$, 
however the rates are smaller by more than an order of magnitude for the 
point chosen for the
study.  Recent analyses of the mass determination of $\stau_{1}$~ and
${\snu}_{\tau_{1}}$~\cite{tata3}  using the continuum production, 
show that with the latter method only an accuracy of $\sim 2\%$ 
for $m_{\stau_1}$ (consistent with the earlier analyses~\cite{MIHO2}) 
and even much worse $6-10 \%$ for $m_{\snu_\tau}$, is possible even after 
a use of optimal polarisation and comparable luninosities as in the 
threshold scan  case. This shows that the threshold scan method will offer 
a better measurement in general. The very high efficiency for 
$\tau$ detection in the TESLA environment might also be playing a role 
in this difference in the accuracies as the other 
analyses~\cite{MIHO1,MIHO2,tata3} use the full Monte Carlo simulation 
using the hadronic decay products of the $\tau$.  However, that does not seem 
to be the full story. While it is true that the threshold scan methods will 
possibly yield more accurate measurement of masses as compared to the
continuum, the low rates for the $\snu_\tau ,\stau_2$ might force one 
to go away from the threshold somewhat, thus sacrificing the accuracy.  
Note also that the branching ratios of the $\snu_\tau$ into different 
channels are not going to be known, a priori. This means that the
normalisation, along with shape will also have to be fitted to the observed 
event rate, which measures cross-section (which we want to measure to 
determine mass) times the branching ratio. While  it is not clear how  much
this will affect the precision with which mass can be extracted, it will
certainly lead to some degradation of its measurement. A preliminary study 
underway~\cite{tata4} to address these issues does not seem to reproduce the 
high accuracy of the mass measurements for $\snu_{\tau}, \snu_\mu$, 
even for the threshold scan.  It is very important to clearly understand
just how well these measurements can be made, as these accuracies 
affect,  crucially, the projected abilities to gleen information about the SUSY 
breaking scale.
 
\begin{figure}[htb]
\centerline{
 \includegraphics*[scale=0.40]{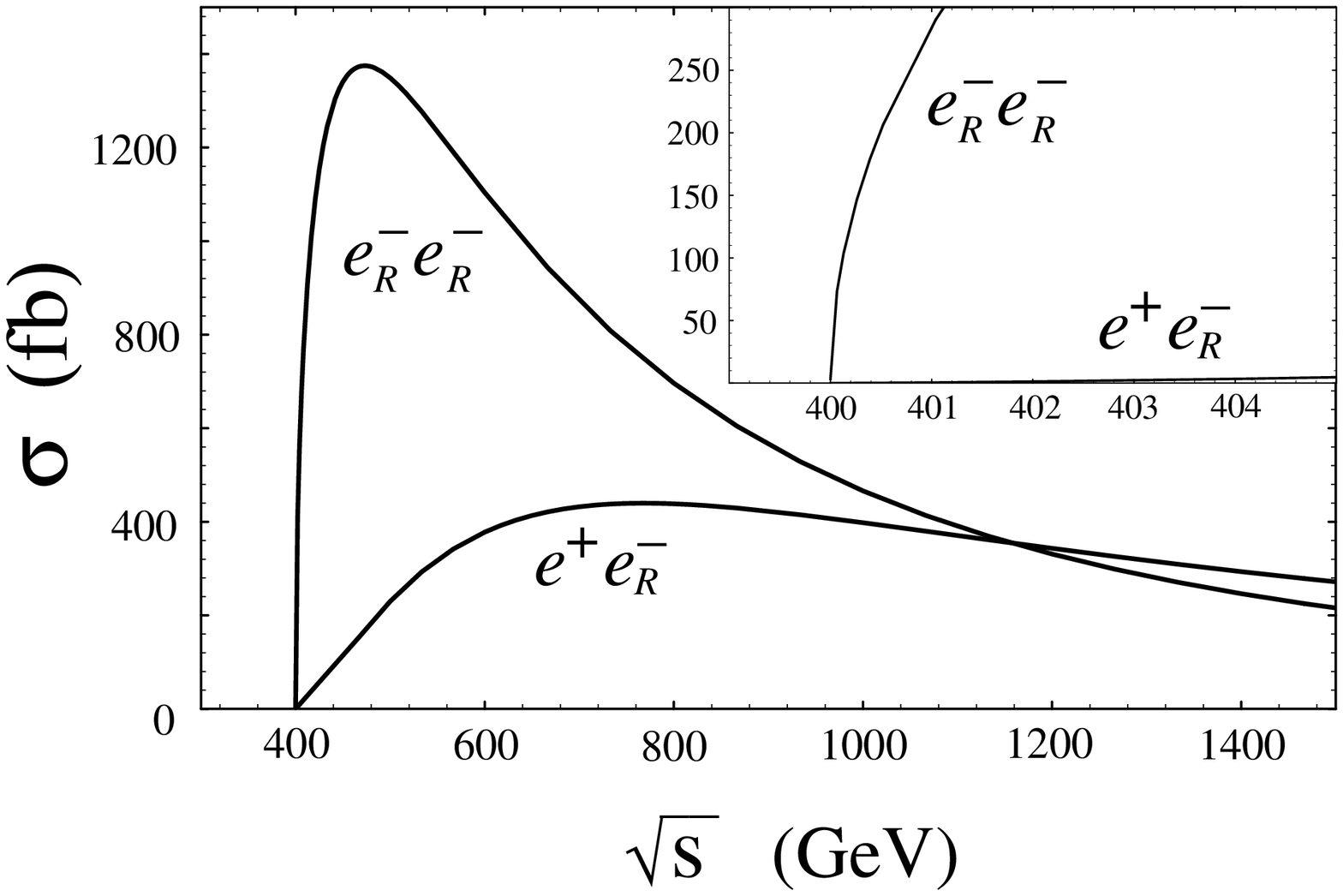}
}
\caption{Different threshold rise of $\sel_R \sel_R^*$ and
$\sel_R \sel_R$ production cross sections for $\epem$ and $\emem$
processes respectively, taken from \protect\cite{fee}.}
\label{F:rgplen:3}
\end{figure}

Using $e^{-}e^{-}$ collisions instead of $e^{+}e^{-}$ gives an interesting
advantage in the study of selectron production~\cite{fee}.
$\sel_{R} \sel_{R}$ production in  $e^{-}e^{-}$ collisions proceeds only
through the $t$ channel diagram as opposed to the case of  $e^{+}e^{-}
\to \sel^{*}_{R} \sel_{R}$ production and  hence has a threshold rise
$\propto \beta $ instead of the  $\beta^{3}$ as in the latter case. The former
makes the study of $\sel_{R}$ production in $e^{-}e^{-}$ collisions
much more sensitive to the nature of $\N0_1$ and the latter
has the potential of increasing the accuracy of the  $\sel_{R}$ mass
determination through threshold scans. This difference in the threshold rise
of the cross-sections in the two cases is shown in Fig.~\ref{F:rgplen:3} 
taken from~\cite{fee}.
However, it must be pointed out that this study does not include effects
of beamstrahlung and ISR. A report presented at this meeting~\cite{heusch} 
shows that these might blur the distinction, at least  for the 
X-band designs.  Also it should be remembered that selectrons are the
only sparticles that  can be produced at an $e^-e^-$ collider.

The subpermille achievable accuracy for sparticle mass measurements that 
the analysis in  Ref.~\cite{MB1} (cf. Table~\ref{T:rgplen:2}) seems to
indicate by the threshold scan method, underlies the
need of the study of higher order effects in all the studies and that has 
become the state of the art of theoretical calculations. Inclusion of effects 
of the finite width of the
smuon~\cite{MB2} or that of higher order corrections to $\Chip_{i} \Chim_{j}$ 
production~\cite{extra9} or the contribution of the nonresonant
production of $\mu^{+} \mu^{-} \N0_{1} \N0_{1} $ ~\cite{freitas} 
on the precision of the mass measurement using threshold scans are being 
studied.

\vspace{0.5cm}
\noindent\underline{{\it Squarks}}

\begin{figure}[htb]
\centerline{
 \includegraphics*[scale=0.40]{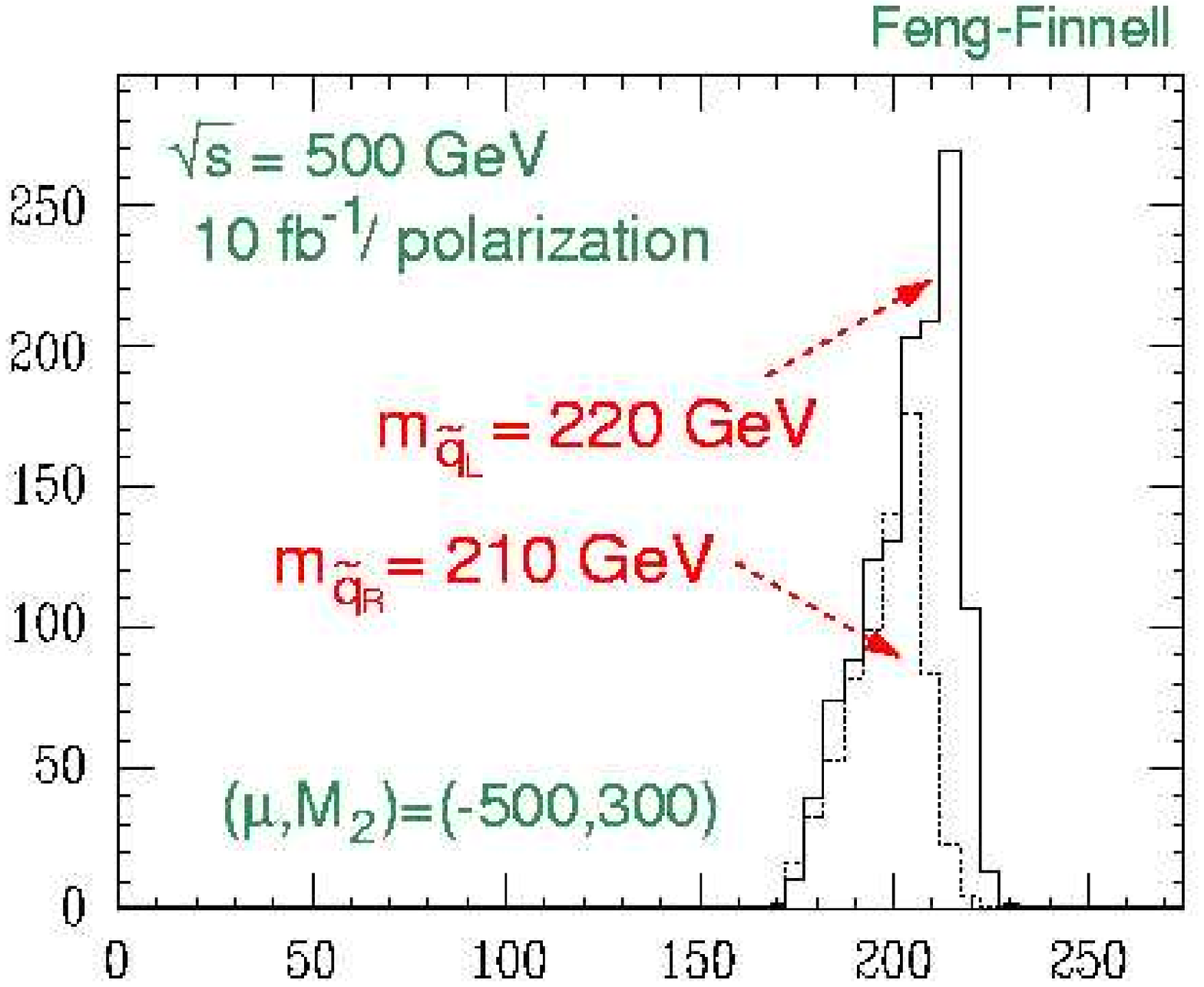}
 \includegraphics*[scale=0.40]{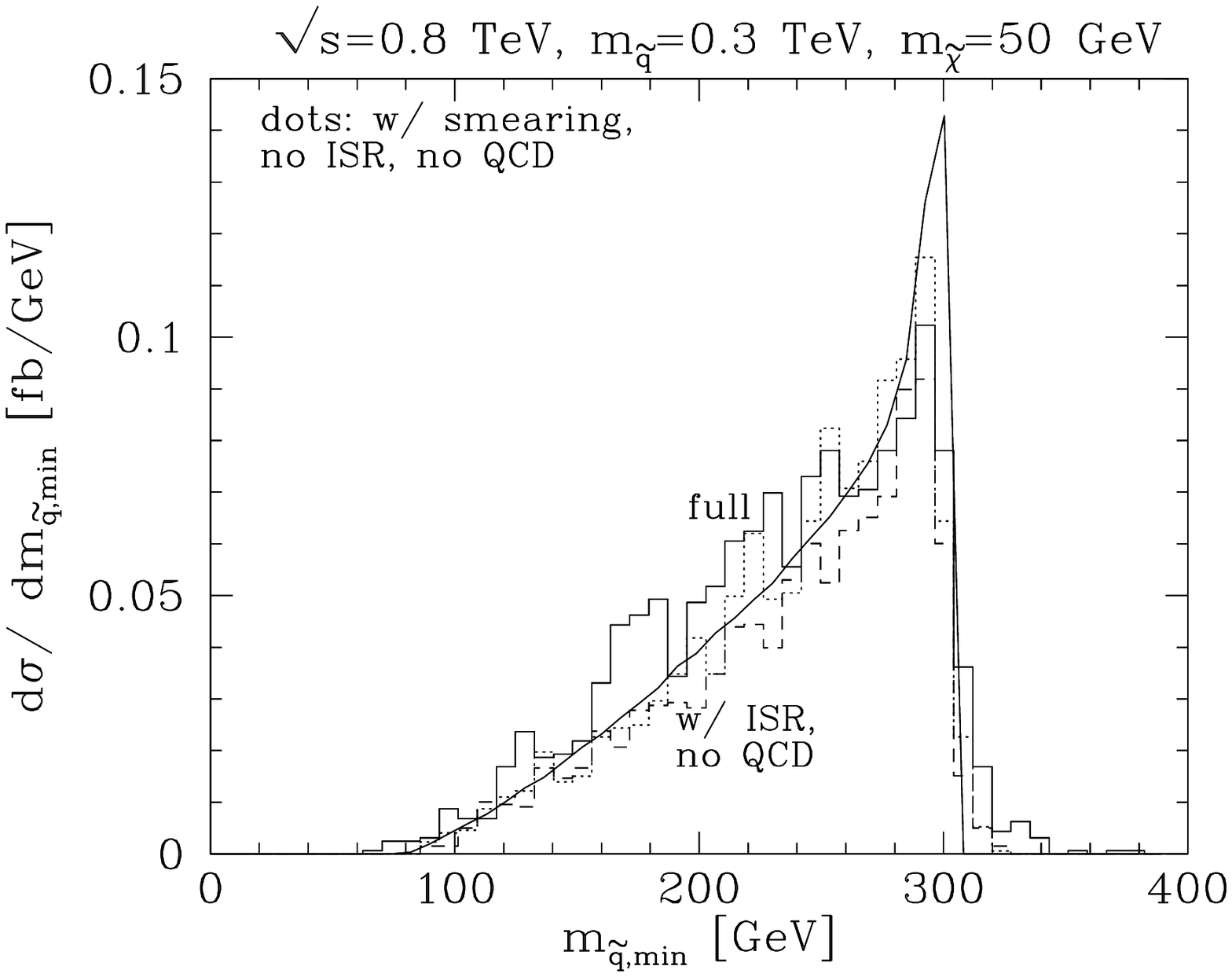}
}
\caption{Accuracy of determination of $m_{\sq}$ using the $m_{\sq,min}$ defined 
in~\protect\cite{FF}.}
\label{F:rgplen:5}
\end{figure}

\noindent Clearly squarks are the only strongly interacting particles about
whom direct information can be obtained at the  $e^{+}e^{-}$ collider. For
the strongly interacting sfermions (squarks) the decay is $\sq \to
q \N0_{1}$. As a result, one has to study the end point of the
distribution in $E_{jet}$. The hadronization effects can in principle 
deteriorate
the accuracy of the determination of $m_{\sq}$. An alternate estimator
~\cite{FF} of  $m_{\tilde{q}}$ is the peak of the distribution in the minimum
kinematically allowed mass of the $q\N0_{1}$ system produced in
\begin{figure}[ht]
\centerline{
 \includegraphics*[scale=0.4]{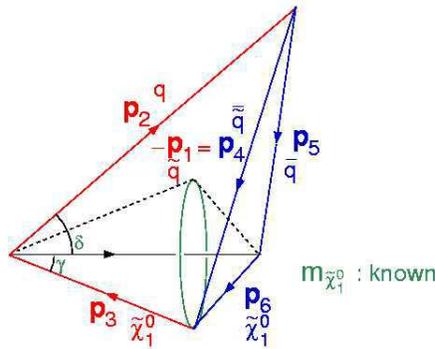}
}
\caption{Determination of the minimum kinematically allowed squark mass, 
following Ref.~\protect\cite{FF}.}
\label{F:rgplen:4}
\end{figure} 
$\tilde{q}$ decay; $m_{\tilde{q}, min}$. The minimum squark mass corresponds 
to maximum possible $|\vec p_4|$ and can be easily determined following the
construction in Fig.~\ref{F:rgplen:4}. The figure in the left panel of
Fig.~\ref{F:rgplen:5}, taken from Ref.~\cite{FF} shows the efficacy of 
this estimator for a 500 GeV machine with $10$ fb$^{-1}$ luminosity per 
polarisation, the latter 
being used for separating $\sq_{L}/ \sq_{R}$ contributions, for
a particular point in the MSSM parameter space. Figure in the right panel
shows that this variable provides a good estimate of  $m_{\sq}$ even
after radiative corrections, both in production and decay, have been
included~\cite{DEKG}.

If the squarks are lighter than the glunios, any information on gluino 
masses at an  $e^{+}e^{-}$ collider can only come from the assumed relations 
between the masses of the electroweak and strong gauginos.

\subsection*{Precision determination of mixings}
The mixing between various interaction eigenstates in the gaugino sector as
well as the, in general, large mixing in the L-R sector for the third
generation squarks and sleptons, is decided respectively by $M_1, M_2, \mu,
\tan\beta$ and $\mu, \tan\beta, A $ as well as various scalar mass
parameters. So clearly an accurate measurement of these mixings along with
the precision measurements of masses offers further clues to physics at
high scale. Table~\ref{T:rgplen:2} shows in the last column the 
parameters whose values can be extracted from mass measurements of various 
EW sparticles; the sleptons and the chargino/neutralinos.

\begin{figure}[htb]
\centerline{
 \includegraphics*[scale=0.80]{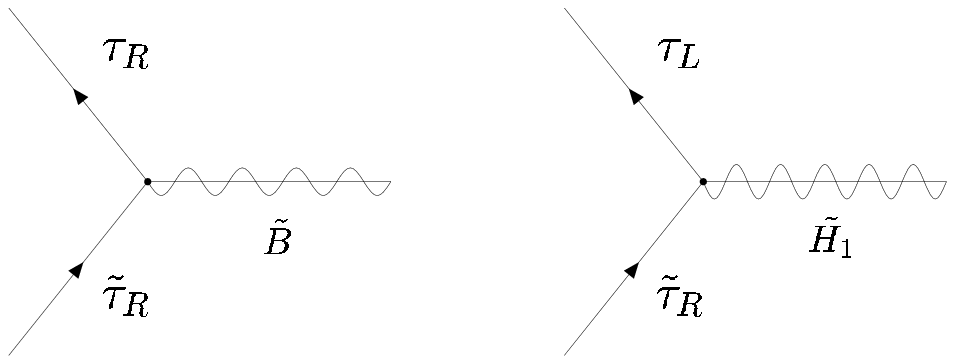}
}
\caption{ $\stau_{R} \to \tau \N0_{1}$. }
\label{F:rgplen:6}
\end{figure}

Possibilities of the determination of L-R mixing in the third generation
sfermions have been investigated~\cite{MIHO1,MIHO2,SM}. The mass eigenstates 
can be written
down in terms of the interaction eigenstates for,\eg, staus as
$\stau_{1} = \stau_{L} \cos \theta_{\tau} + \stau_{R}
\sin \theta_{\tau}$,$\stau_{2} = \stau_{L} \sin \theta_{\tau}
+ \stau_{R} \cos\theta_{\tau}$.
It is clear that polarised $e^{-}$/$e^{+}$ beams can play a crucial role in
determining  $\theta_{\tau}$. Let us, for example, consider
$e^{+} e^{-} \to \stau_{1} \stau_{1}^{*}$. Further
let us consider the case of 100 \% polarisation in particular. The pair
production proceeds through an exchange of $\gamma$/Z in $s$-channel. For
energies $\sqrt{s}$ $>$ $>$ $m_{Z}$, with $P_{e^{-}}$ = 1, one can
essentially interpret this $s$-channel exchange of  $\gamma$/Z as an U(1)
gauge boson, $B$.  In this limit $\sigma (\stau_{R})$ = 4
$\sigma (\stau_{L})$. Thus it is clear that a measurement of $\sigma
( e^{+} e^{-} \to  \stau_{1}^{*}  \stau_{1} )$ along with a
knowledge of polarisation of $e^{-}$ beam can lead to an extraction of $\cos
\theta_{\tau}$. Further the polarisation of $\tau$ produced in
$\stau_{1}$ decay provides a measurement of the mixing angle in the
neutralino sector as well. Let us consider $\stau_{R} \to \tau
\N0_{1}$ depicted in Fig.~\ref{F:rgplen:6}. The $\tilde{B}$
component of $\N0_{1}$ produces  $\stau_{R}$, whereas
the higgsino
component will flip the chirality and produce $\stau_{L}$. Thus the
measurement of $ \stau_{1}^{*}  \stau_{1}$ production with
polarised $e^{-}$ beams and the polarisation of decay $\tau^{s}$ can 
give very useful information on both the mixings: the $L-R$ mixing in the
stau sector and the mixing in the neutralino sector. The $\tau$ polarisation 
can be measured by looking at the energy distribution of the decay 
product $\rho$ in the hadronic decay of $\tau$~\cite{MIHO1}.
\begin{figure}[ht]
\centerline{
 \includegraphics*[scale=0.35]{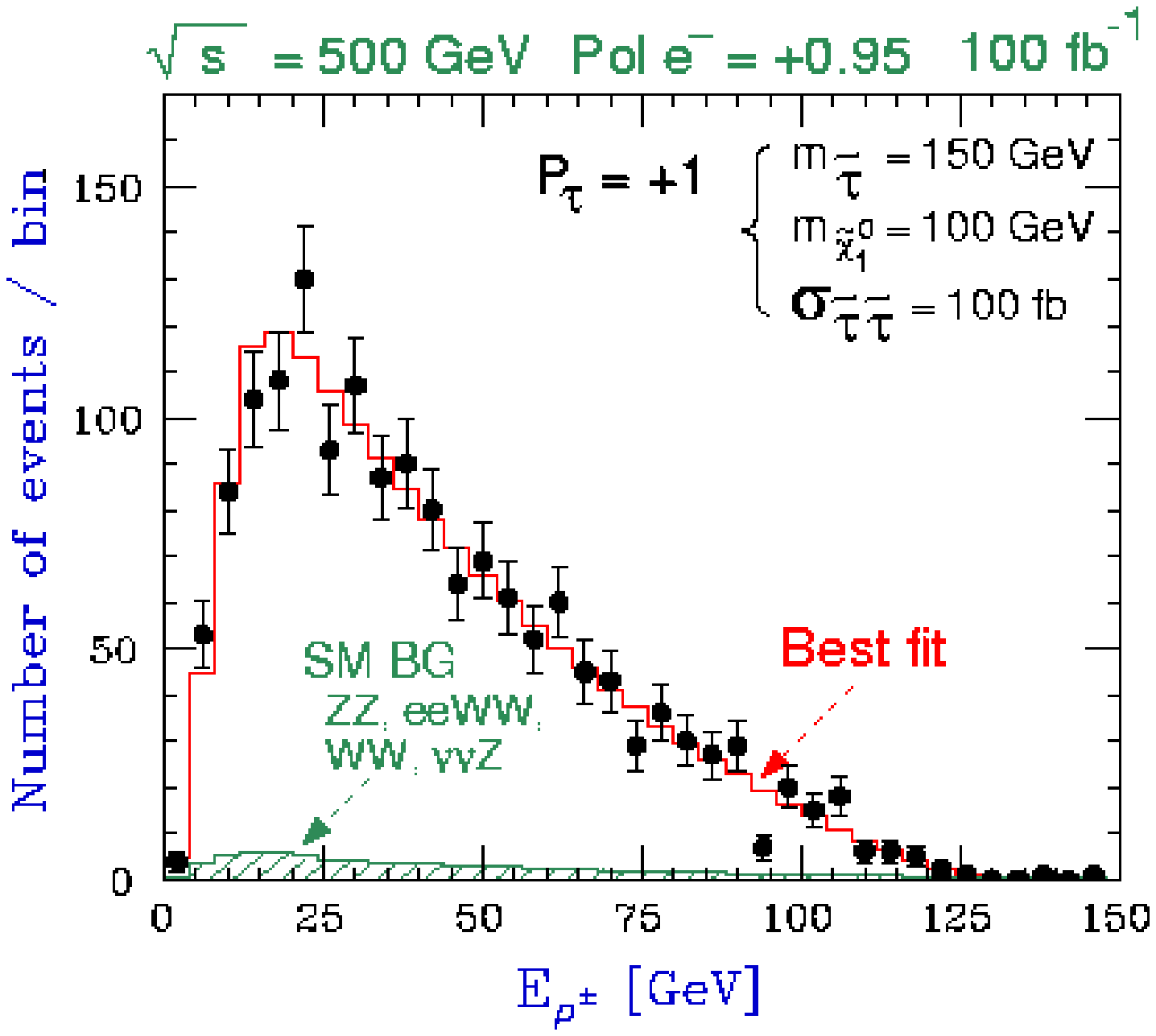}
 \includegraphics*[scale=0.35]{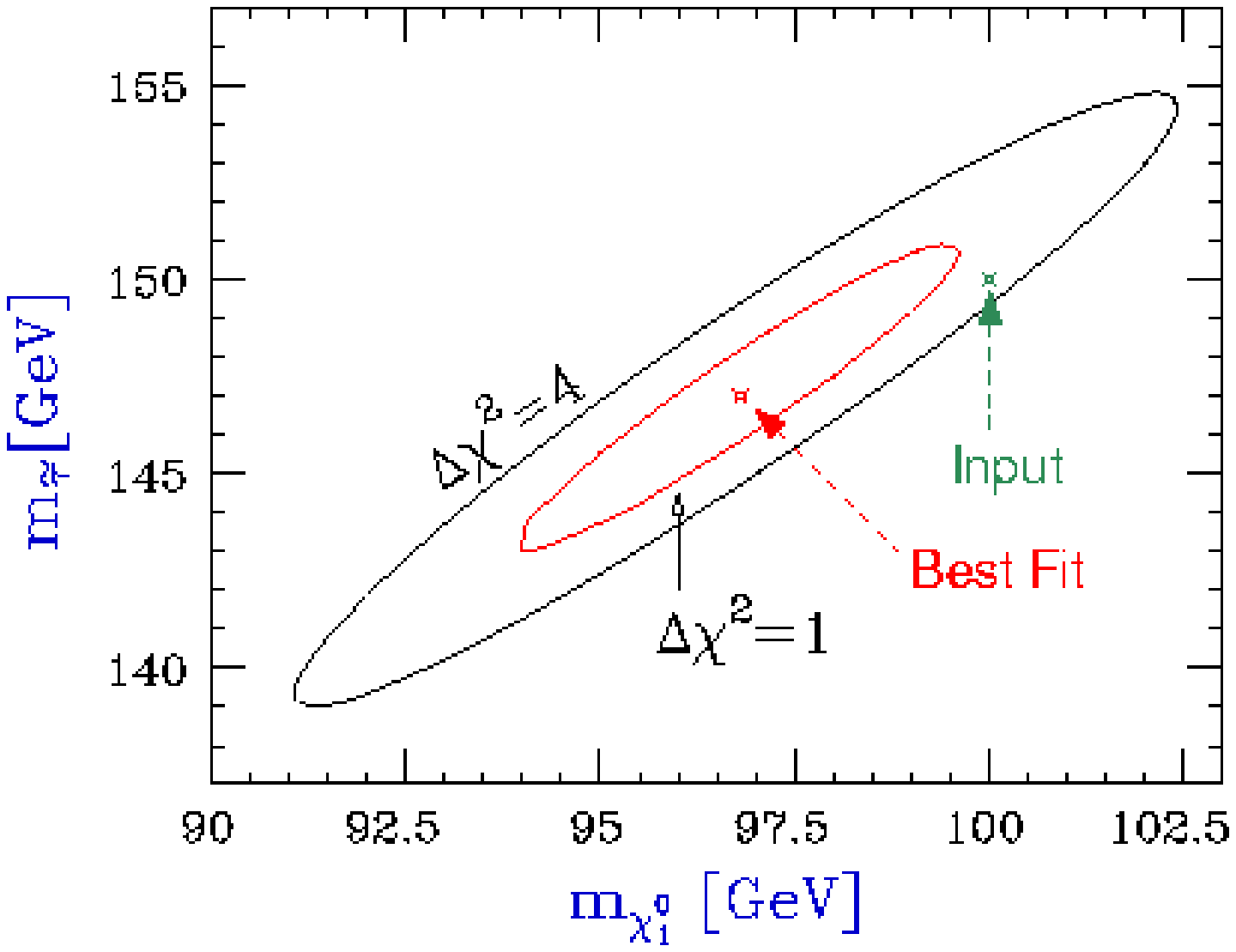}
}
\caption{Precision of the determination of $m_{\stau_{1}}$ and
$m_{\N0_{1}}$ for luminosities and polarisation 
mentioned in the figure. Taken from Ref.\protect\cite{MIHO2}.}
\label{F:rgplen:7}
\end{figure} 
Fig.~\ref{F:rgplen:7}~\cite{MIHO2} shows the possible accuracy of a
simultaneous determination of $m_{\tilde{\tau_{1}}}$ -
$m_{\tilde{\chi_{1}^{0}}}$ from the determination of the end points of the
energy spectrum, for $\int {\cal L} dt = 100\ {\rm fb}^{-1},P_{e^{-}} = 0.95$ 
and $\sqrt{s} = 500$ GeV. The input value lies outside the $\Delta
\chi^{2}$ = 1 contour around the best fit value. However, if
$m_{\N0_{1}}$ is assumed to be known, then  $\Delta m_{\stau_{1}}$ goes 
down considerably and a 1-2\% determination at 
\begin{figure}[htb]
\centerline{
 \includegraphics*[scale=0.35]{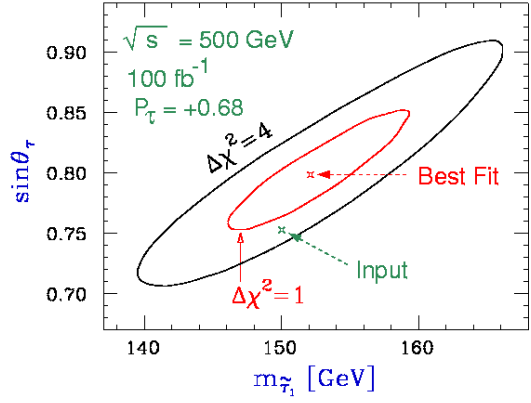}
}
\caption{Simultaneous determination of $\sin\theta_\tau$ and $m_{\stau_{1}}$, 
taken from Ref.\protect\cite{MIHO2}.}
\label{F:rgplen:7p}
\end{figure} 
$1 \sigma$ level is possible. Fig.~\ref{F:rgplen:7p} shows accuracy
of sin$\theta_{\tau}$ determination for the same choice of parameters and we
see that  $\Delta (\sin \theta_{\tau}) < 0.03 $.
\begin{figure}[htb]
\centerline{
 \includegraphics*[scale=0.35]{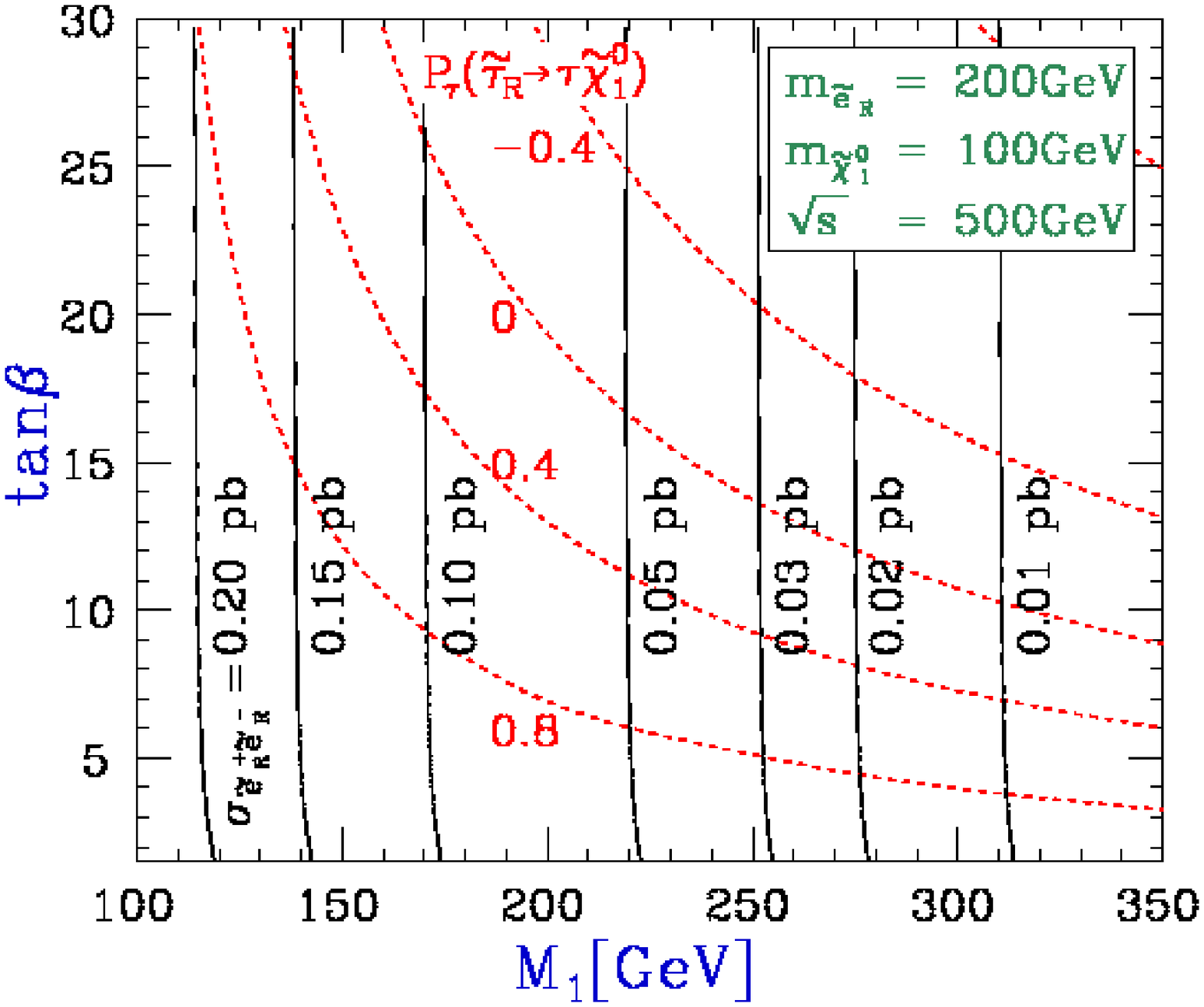}
 \includegraphics*[scale=0.35]{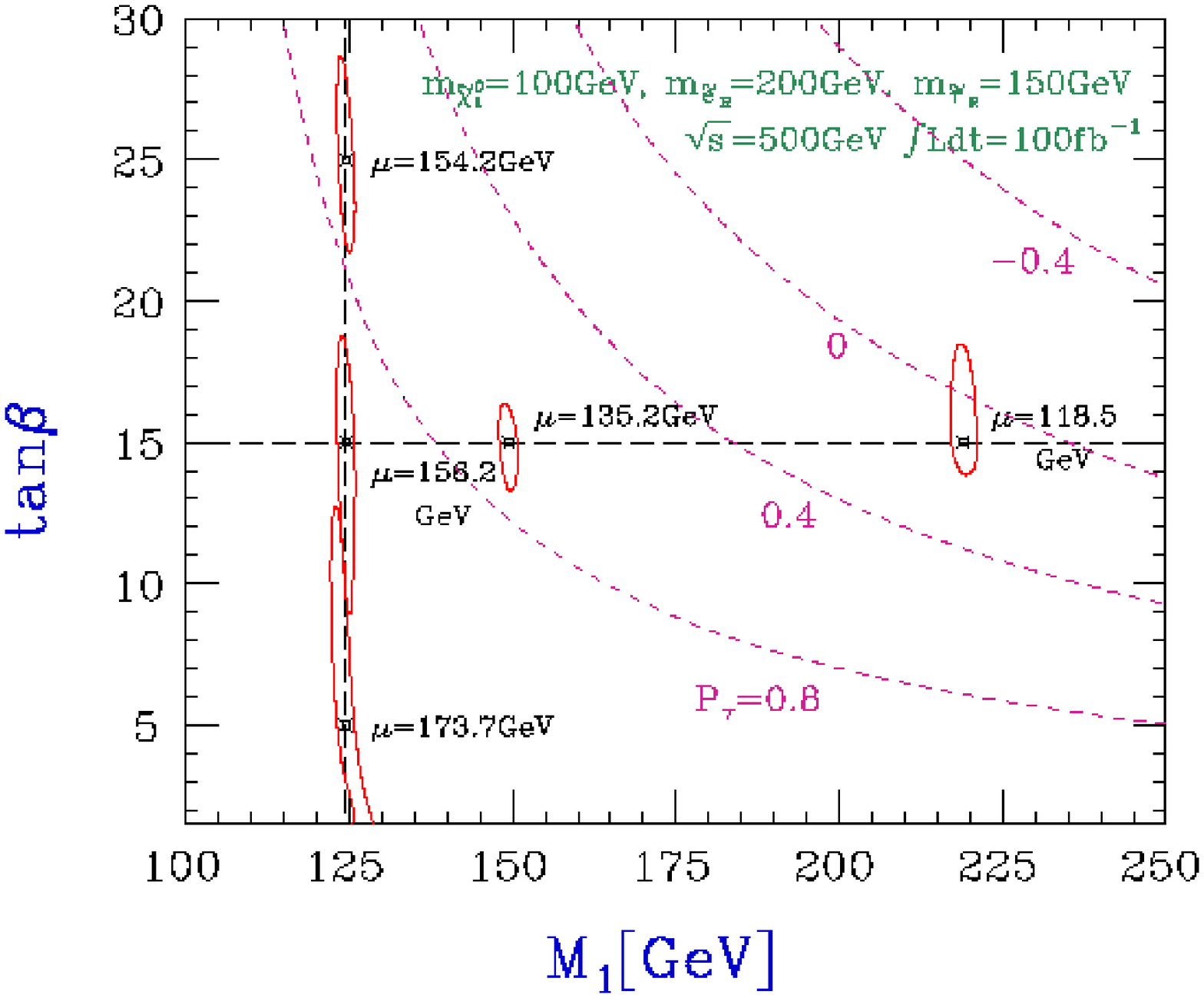}
}
\caption{ Contours of constant cross-section in $M_{1}$ - tan$\beta$ plane 
and accuracy of tan$\beta$ measurement, taken from Ref.~\protect\cite{MIHO2}.}
\label{F:rgplen:8}
\end{figure} 
Further, Fig.~\ref{F:rgplen:8}~\cite{MIHO2} shows in the left panel 
the contours of constant cross-section  
$\sigma ( e^{+} e^{-} \to  \sel_{R}^{*} \sel_{R} )$, and of
constant polarisation  $P_{\tau}$ ( $\stau_{R} \to \tau \tilde{\N0_{1}}$) 
in the $M_{1}$ - $\tan\beta$ plane (this analysis
assumes universal gaugino masses at a high scale). The figure in the right
panel shows the accuracy one can expect from a simultaneous study of 
these two measurements. It shows that the method has potential of 
a good $\tan\beta$ determination at large $\tan\beta$.

\begin{figure}[htb]
\centerline{
 \includegraphics*[scale=0.75]{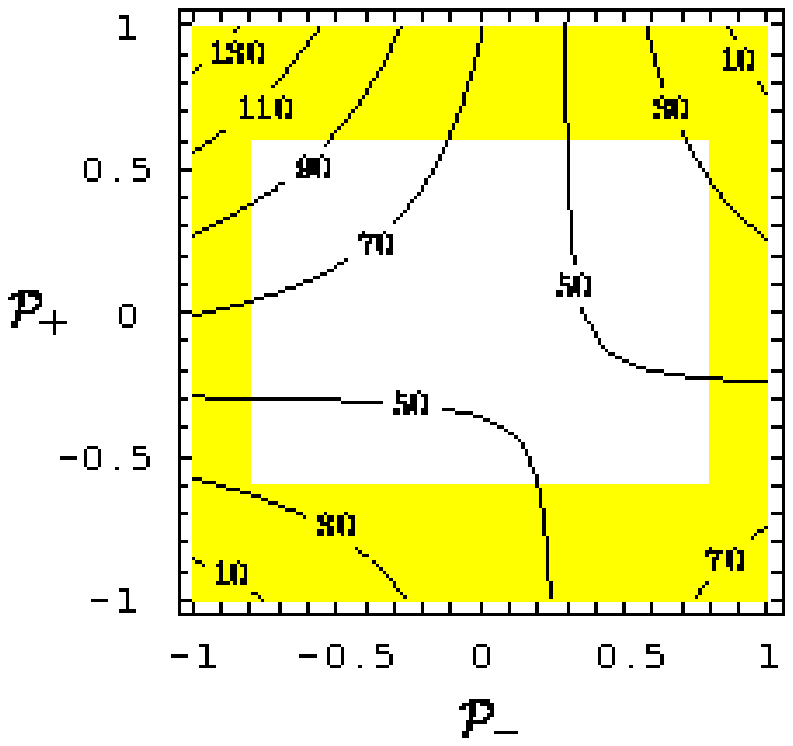}
 \includegraphics*[scale=0.25]{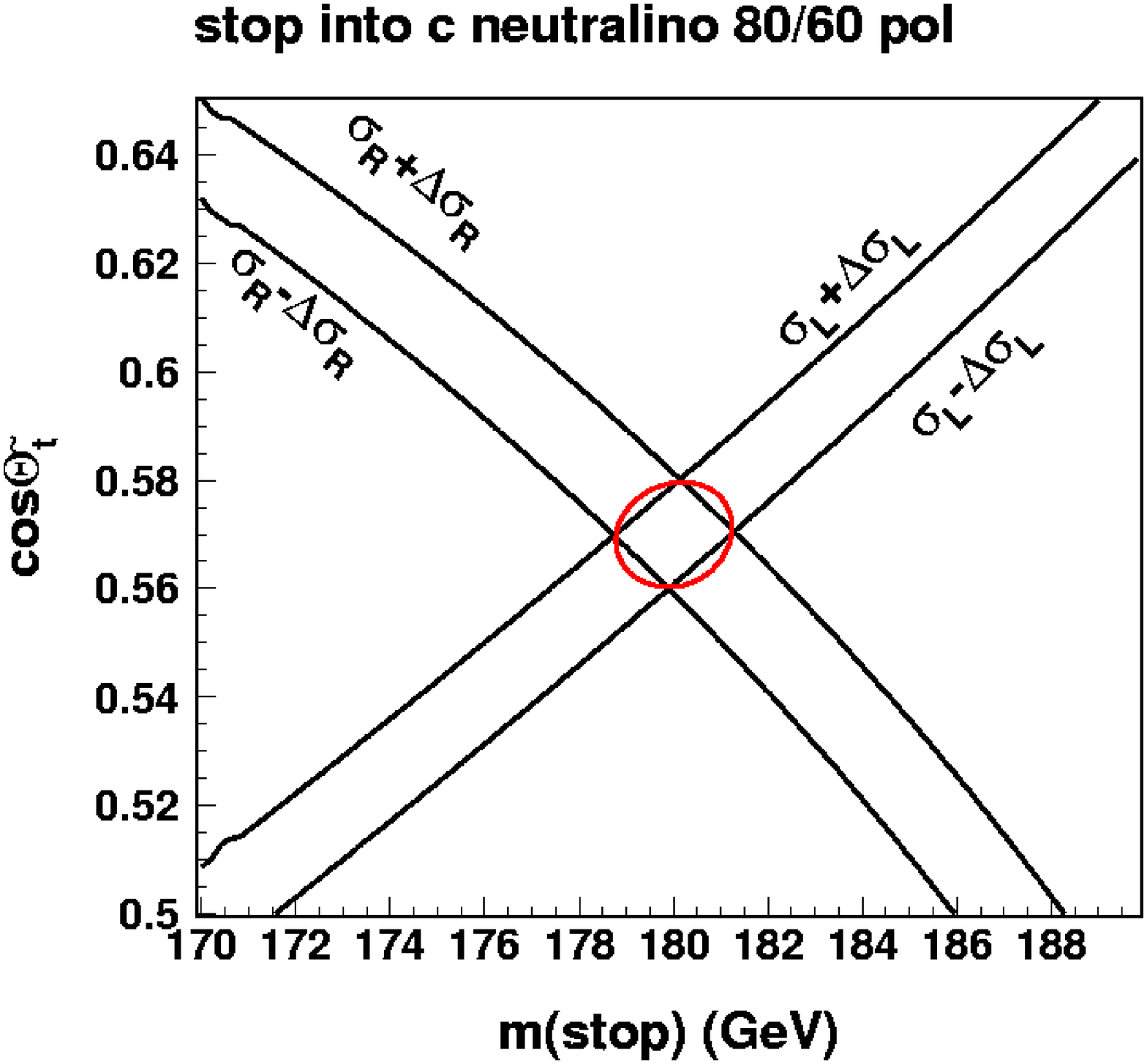}
}
\caption{Polarisation dependence of the production cross-section for
$\st_1 \st_1^*$ pair and the expected accuracy of a simultaneous
determination of $\cos\theta_t$ and $m_{\st_1}$ at the TESLA collider.}
\label{F:rgplen:9}
\end{figure}

Left panel in Fig.~\ref{F:rgplen:9}~\cite{SM,teslatdr} shows contours 
of constant cross-section for
$\st_{1} \st_{1}^{*}$ production at TESLA for
$m_{\st_{1}} = 180 $ GeV and $\cos\theta_t = 0.57$ as a function of the
polarisation $P_{-}$/$P_{+}$ of the $e^{-}$/$e^{+}$ beam. This shows
how it is important to have polarisation for {\bf both} beams. The right
panel~\cite{stop_mc,teslatdr} shows also the accuracy of $\cos\theta_{t}$ -
$m_{\st}$ measurement using $\st_{1} \to c \N0_{1}$
decay mode for stops and 80/60 polarisation for $e^{-}$/$e^{+}$ beam. The
figure shows that at TESLA one can reach  $\Delta (\cos\theta_{t}) = 0.003,
\Delta (m_{\st}) = 0.08$. It is interesting to note that if $m_{\st_1}$ is
in the mass range of 180-225 GeV , the $2 \gamma$ decay mode of the 
lightest higgs may not be accessible at the LHC and even the visibility 
of the $\st_1$ in this mass range at the LHC has not been completely
analysed.

\begin{figure}[htb]
\centerline{
 \includegraphics*[scale=0.35]{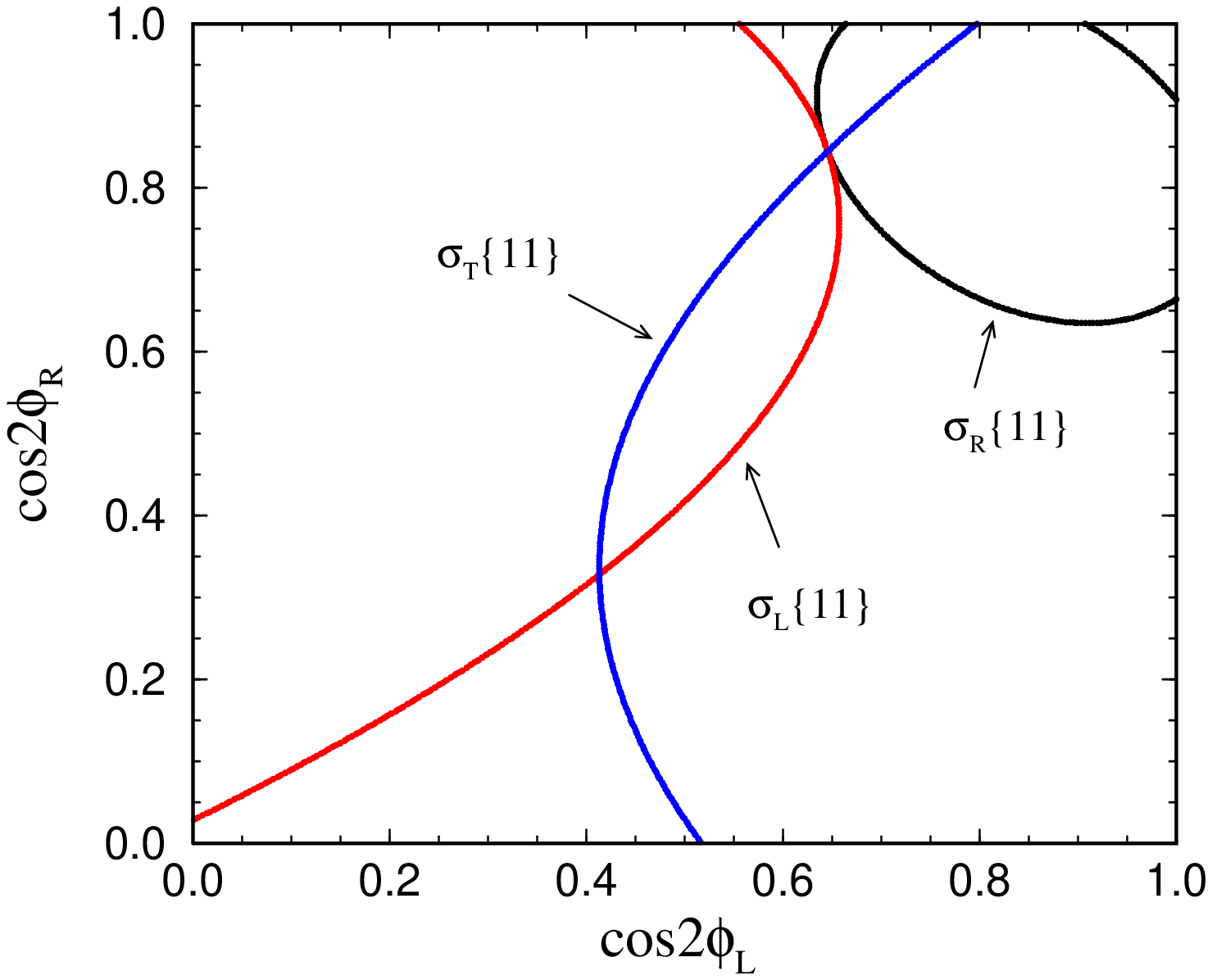}
}
\vspace {0.20in}
\centerline{
 \includegraphics*[scale=0.50]{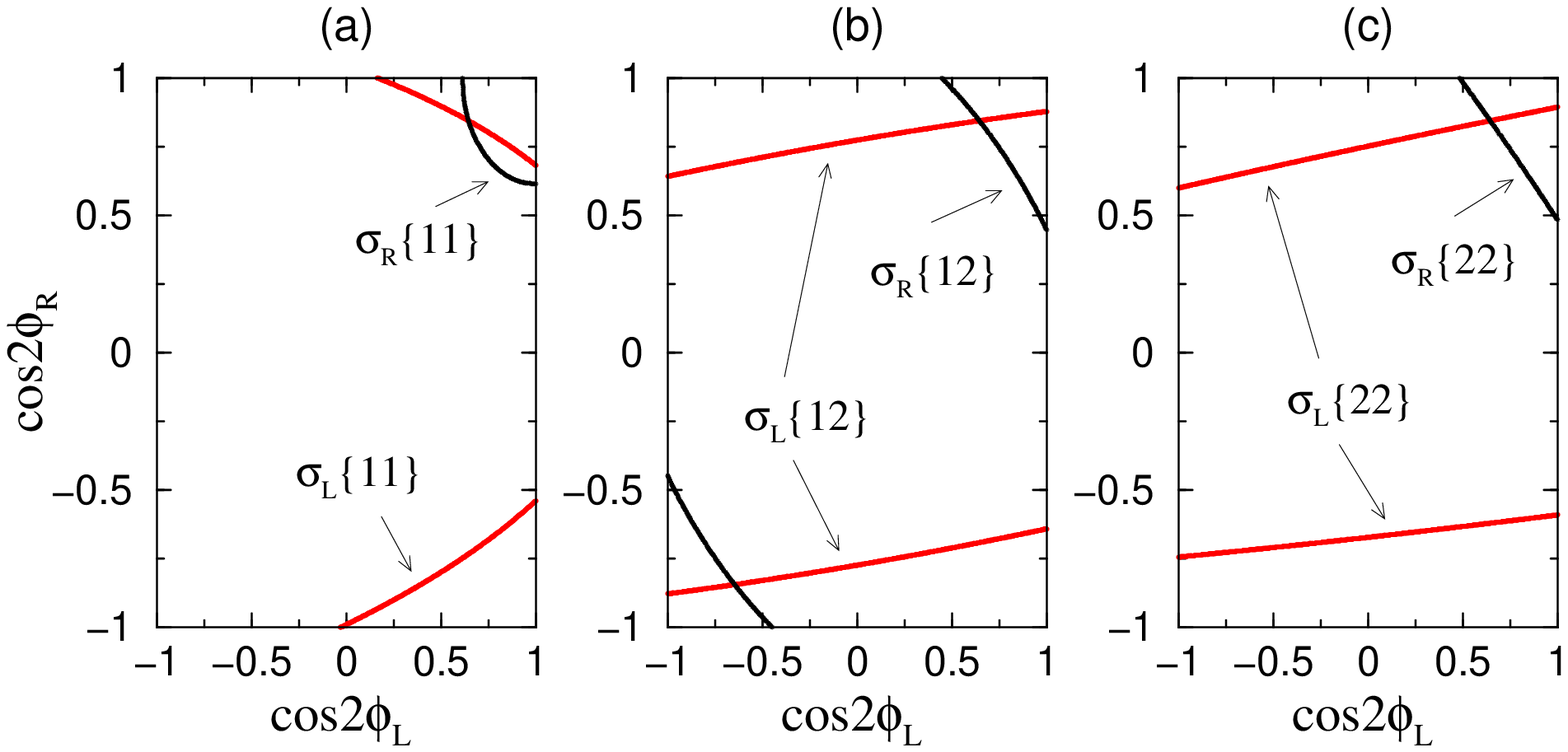}
}
\caption{Demonstration of unique determination of mixing angles in the
chargino sector using polarised beams~\protect\cite{Z1,Z2,Z3}.}
\label{F:rgplen:10}
\end{figure}

A study of the chargino sector at the LC can provide a precision determination
of the higgsino-gaugino mixing and consequently an accurate determination
of {\bf all} the Lagrangian parameters which dictate the properties 
of the chargino sector. This requires, along with  the determination of 
$m_{\Chipm_1}, m_{\N0_1}$ and $\sigma^{\rm tot}_{\Chip_1,\Chim_1}$, 
a study of either the dependence of the production cross-section on 
the initial beam polarisations or the polarisation of the produced charginos 
through the angular distribution of their decay products. Since 
$\sigma(\epem \to \Chim_i \Chip_j)$ depends on  $m_{\snu}$,
its knowledge is also necessary. This 
can be obtained by studying the energy dependence of the $\sigma^{\rm tot}$,
even if $m_{\snu}$ is beyond the kinematic range of the collider. If only 
the lightest chargino is available kinematically, then one can determine the
mixing angles in the chargino sector $\Phi_L, \Phi_R$ defined through 
\be
\Chim_{1L} = \cos \Phi_L \tilde W^{-}_{L} + \sin \Phi_L \tilde H^-_{2L},
\Chim_{1R} = \cos \Phi_R \tilde W^-_R + \sin \Phi_R \tilde H^-_{1R},
\ee
only upto a two fold ambiguity.  However, this can be removed, 
using the information on the transverse polarisation, as shown in               
Fig.~\ref{F:rgplen:10}. If both the charginos are accessible kinematically,
$\cos 2\Phi_R, \cos 2\Phi_L$ can be determined uniquely through
measurements of $\sigma_{L/R} (\Chip_i \Chim_j)$, as shown in the lower panel 
of the same figure. It has been shown, in a purely theoretical study
~\cite{Z1,Z2}, in the context of TESLA, using only statistical 
errors, that with
$\int {\cal L} dt = 1\ ab^{-1}, \cos 2\Phi_L, \cos 2\Phi_R$ can be determined
to an accuracy of $\sim 1-3 \% $. Along with the information on $m_{\Chip_i}$
the mixing angles can then lead to an unambiguious determination of the
Lagrangian parameters $M_2,\mu$ and $\tan \beta$. However, since all the
variables are proportional only to $\cos 2\beta$, the accuracy of $\tan \beta$
determination is rather poor at high $\tan \beta$. At high $\tan \beta$,
measurements in the slepton sector (stau/selectron) discussed 
earlier~\cite{MIHO2} afford a better determination. Alternative ways of 
extracting $\tan\beta$ from a study of the Higgs sector~~\cite{feng_h,han} 
have been suggested.  But of course these
need access to the heavier higgses $A,H$ and $H^\pm $. In view of the current
LEP limits on the $\tan \beta - m_A$ plane, indications are that such
determination might require $\sqrt{s}$ values larger than the $500$ GeV that is
envisaged in the first stage for an LC. A better handle on the $\tan \beta$ in
the large $\tan \beta$ range is offered by the studies of the $H/A, H^{\pm}$ 
sector at the LHC~\cite{atlas_tdr}.

Similar studies of the neutralino sector~\cite{MGP1,kneur1,kneur} show that 
one can extract $\tan \beta, M_2, \mu$ as well as the relative phase of 
$\mu/M_2$~\cite{kneur} in case it is nonzero. Use of polarisation for both
the beams~\cite{gtalk} allows extraction of $M_1,M_2,\mu $ and $\tan \beta$, 
{\it without} assuming the unification relation among $M_1,M_2$~\cite{MGP1}. 
Availability of polarisation of {\it both} the beams seems to increase the 
accuracy of the measurement substantially. The sensitivity to the departure 
from the universal gaugino masses increases further~\cite{wuerz}  by using the $e \gamma$ 
option of the collider should that be realised.

\subsection*{Scenarios other than mSUGRA and MSSM}
\noindent \underline{{\it AMSB}}

Another interesting set of studies of the chargino/slepton sector is in the 
context of the AMSB models, wherein one expects an almost degenerate pair
of the lightest neutralino/chargino which are essentially winos. Since the mass 
difference is expected to be $165~{\rm MeV} < \Delta M <1 $ GeV,  
$\Chip_1$ has $\sim 96 \% $ B.R.  in the $\pi^+ \N0_1$ channel. 
Depending on the mass-difference $m_{\Chip_1} - m_{\N0_1}$ and hence the 
life time of $\Chip_1$, the signature can be
either a high momentum track stoping in the vertex detector or a displaced
vertex which can be inferred from the impact parameter of the soft  decay pion.
Feasibility of studying the pair production of the charginos~~\cite{gm} 
as well as the left selectrons(smuons)~\cite{dsp}, at the NLC in 
this scenario has been demonstrated.  The study~\cite{gm} shows that 
using different techniques, it is possible to probe the chargino/neutralino 
masses right upto the kinematic limit even in this case.

\vspace{0.5cm}
\noindent \underline{\it {Unstable $\N0_1$}}

\noindent
As discussed in the introduction, the $\N0_1$ is not necessarily stable if
$R$ parity is broken or gravitino is the LSP. In the case of \rpv\ 
apart from the very clean and striking signals due to production of
single sparticle resonance through the \rpv\ couplings
there have 
also been beginnings of detailed investigations~\cite{grr} of possibility of
studying the sparticle signals at the LC in this scenario, when the \rpv\
couplings are small and hence the effect of \rpv\ is seen only in the 
decay of the $\N0_1$. Indeed, the decays of the lightest neutralino give 
rise to significant and striking signals which can be studied with ease
at an LC,  for the case of lepton number violating
$\l$ and $\lp$ couplings. Interestingly, even in the case of the \bv\ $\lpp$
couplings, it seems possible, not only to see the signal due to the 
gaugino/higgsino  production, but also to get information on  the mass of 
the lightest neutralino in this case, which would be particularly  difficult 
for the hadronic colliders to measure. 

\begin{figure}[htb]
\centerline{
 \includegraphics*[scale=0.35]{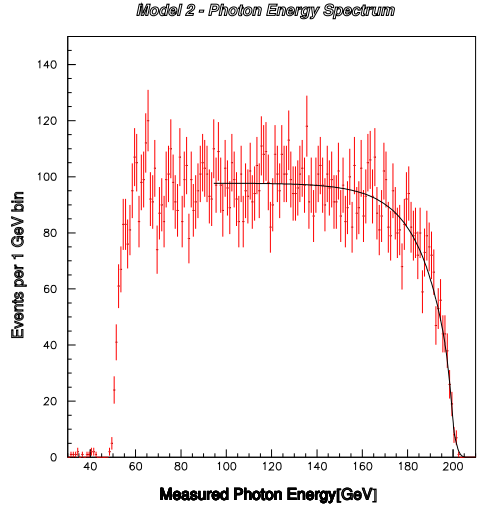}
}
\caption{Accuracy of mass of $\N0_1$  at an LC for GMSB models which have 
the  $\N0_1$ as the NLSP, taken from Ref.\protect\cite{AB}.}
\label{F:rgplen:11}
\end{figure}

In the case of GMSB models, the difference in the search strategy of the
sparticles indeed comes only from the decay of the $\N0_1$ (or for that matter
that of the NLSP). In the literature invesitgations exist for the case where
$\N0_1$ is the NLSP~\cite{AB} as well as 
when $\stau_1$ is the NLSP~\cite{HH}. The former
study includes very detailed analysis of the signals for a decaying
$\N0_1$, for a wide range of GMSB models.  The dominant decay of the
neutralino NLSP is always into a $\tilde G \gamma$ channel. If the lifetime 
of the NLSP is large, pair production of $\N0_1$ and their decay,
can give rise to one or more nonpointing photons, due to the delayed decay of
the NLSP. For larger masses of the NLSP $(\ge 100$ GeV) two body decay into a 
$Z_0 \gamma$ channel followed by the $Z_0$ decay into a $f \bar f$ pair, can 
provide a cleaner signature. A measurement of the upper end point of the
energy spectrum of the decay photon from the $\tilde G \gamma$ decay, 
can provide an accurate measurement of the mass of $\N0_1$. For example, 
with an integrated luminosity of $200\; {\rm fb}^{-1}$, for a $\N0_1$ mass
$200$ GeV, a measurement of $\sim 0.2 \% $ accuracy is possible. This is
demonstrated in  Fig.~\ref{F:rgplen:11}. We will see
later, how this can prove very useful in determining the SUSY breaking 
scale in this case.

\section*{Determination of quantum numbers of the sparticles}
\begin{figure}[htb]
\centerline{
 \includegraphics*[scale=0.60]{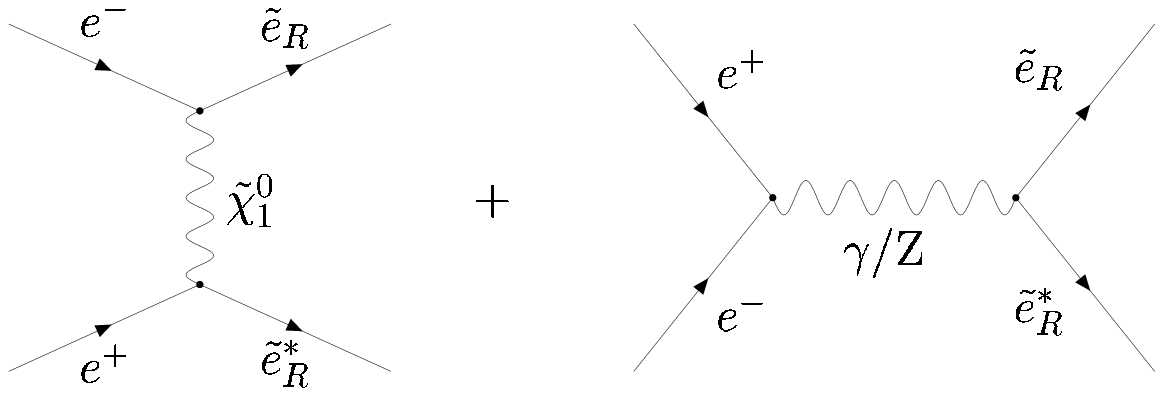}
 \includegraphics*[scale=0.30]{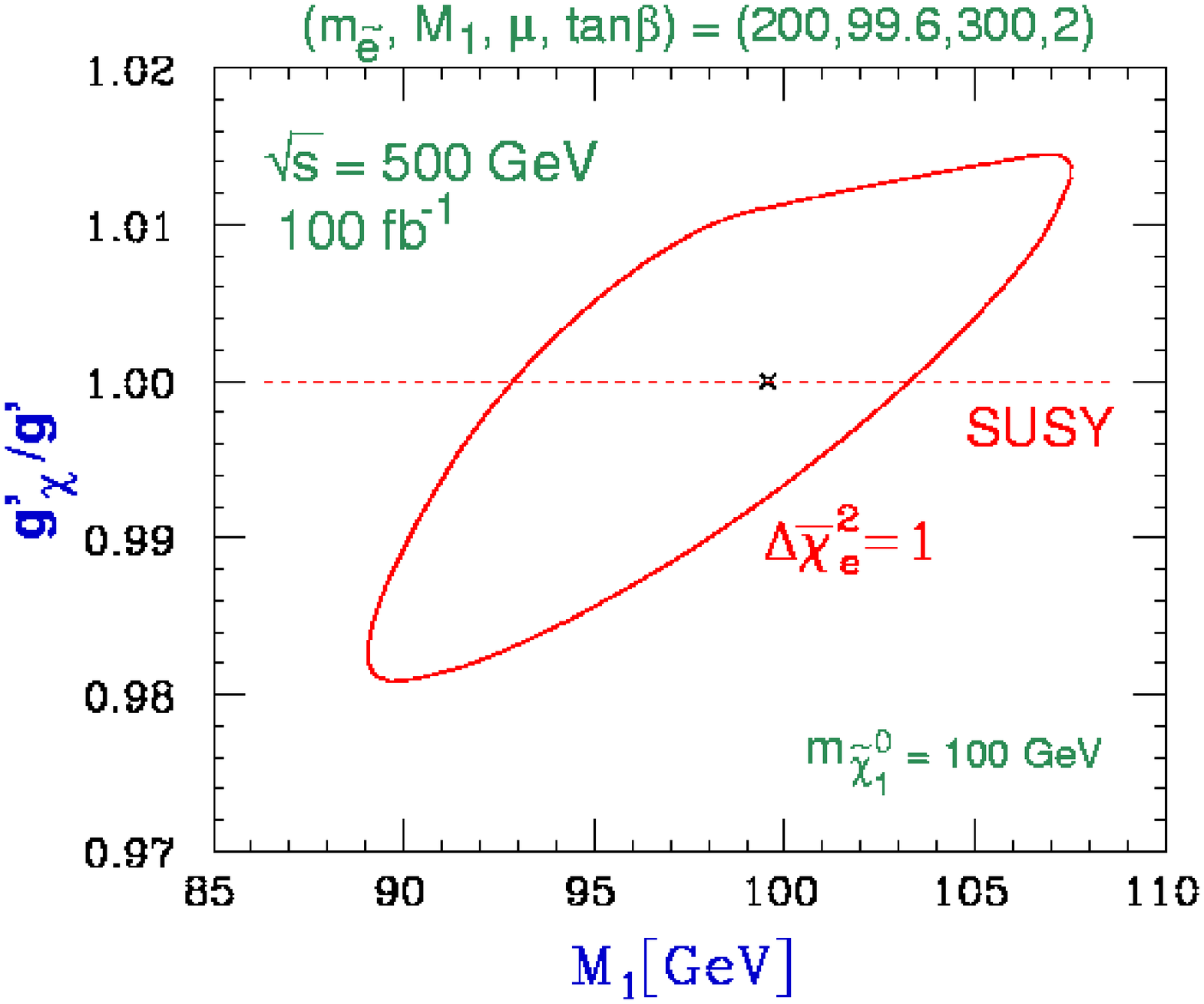}
}

\caption{Simultaneous determination of $M_1$ and $g_{\sel_R e \widetilde B}$,
in a study of $\sel_R \sel_R^*$  production.}
\label{F:rgplen:12}
\end{figure}

Above discussion already shows how an efficient use of polarisation of both
$e^+/ e^-$ beams, allows a high precision determination of the mixings 
among the $L-R$ sfermions as well as in the gaugino-higgsino sector. This is,
indeed, indirectly a determination of the hypercharges of the various 
sparticles. It has been demonstrated~~\cite{FM1}, using realistic 
simulation of the
backgrounds, that it is possible to reconstruct the $\smu$ angular distribution 
in the process $\epem \to   \smu \smu^* \to \mu^+ \mu^-  + \mET $ \  and hence 
determine the spin of the smuon with precision. Further, the cross-section
of $\sel_R \sel_R^* $ production can be used as a very sensitive probe of
the equality of the couplings $g_{\sel e_R \tilde B}$ and $g_{eeB}$. This is 
due the contribution of the $t$ channel diagram shown in the left-hand panel
of Fig.~\ref{F:rgplen:12}, which involves a $\N0_i$ exchange. The contribution 
to the production cross-section of the  $\sel_R \sel_R^*$ pair is
sensitive  to the bino component of $\N0_i$ and hence to the $U(1)$ gaugino 
mass parameter $M_1$ and the coupling $g_{\sel_R e_R B}$. At tree level 
we expect, due to supersymmetry, 
\be
g_{\sel_R e_R \tilde B} = g_{eeB}  = \sqrt{2} g_2 \tan\theta_W =
\sqrt{2} g_1 Y_b = g_Y.
\label{eq:rg:1}
\ee
Using $\sigma (\epem \to \sel_R \sel_R^* ) $ 
and $\frac{d\sigma}{d\cos\theta}
(\epem \to \sel_R \sel_R^*) $, 
one can determine $g_{\sel_R e_R \tilde B}$ and
$M_1$ simultaneously. For an integrated luminosity of $100\ {\rm fb}^{-1}$,
$Y_b$ of Eq.~\ref{eq:rg:1} can be determined to an accuracy of 
$1\%$~\cite{MIHO2}. This is shown in  the right panel of 
Fig.~\ref{F:rgplen:12}. This expected accuracy is actually 
comparable to the size of the SUSY radiative corrections~\cite{MIHO3}
 to the tree
level equality of Eq.~\ref{eq:rg:1} and hence this measurement can serve as
an indirect probe of the mass of the heavy sparticles. We will get to this 
later.
\begin{figure}[htb]
\centerline{
 \includegraphics*[scale=0.35]{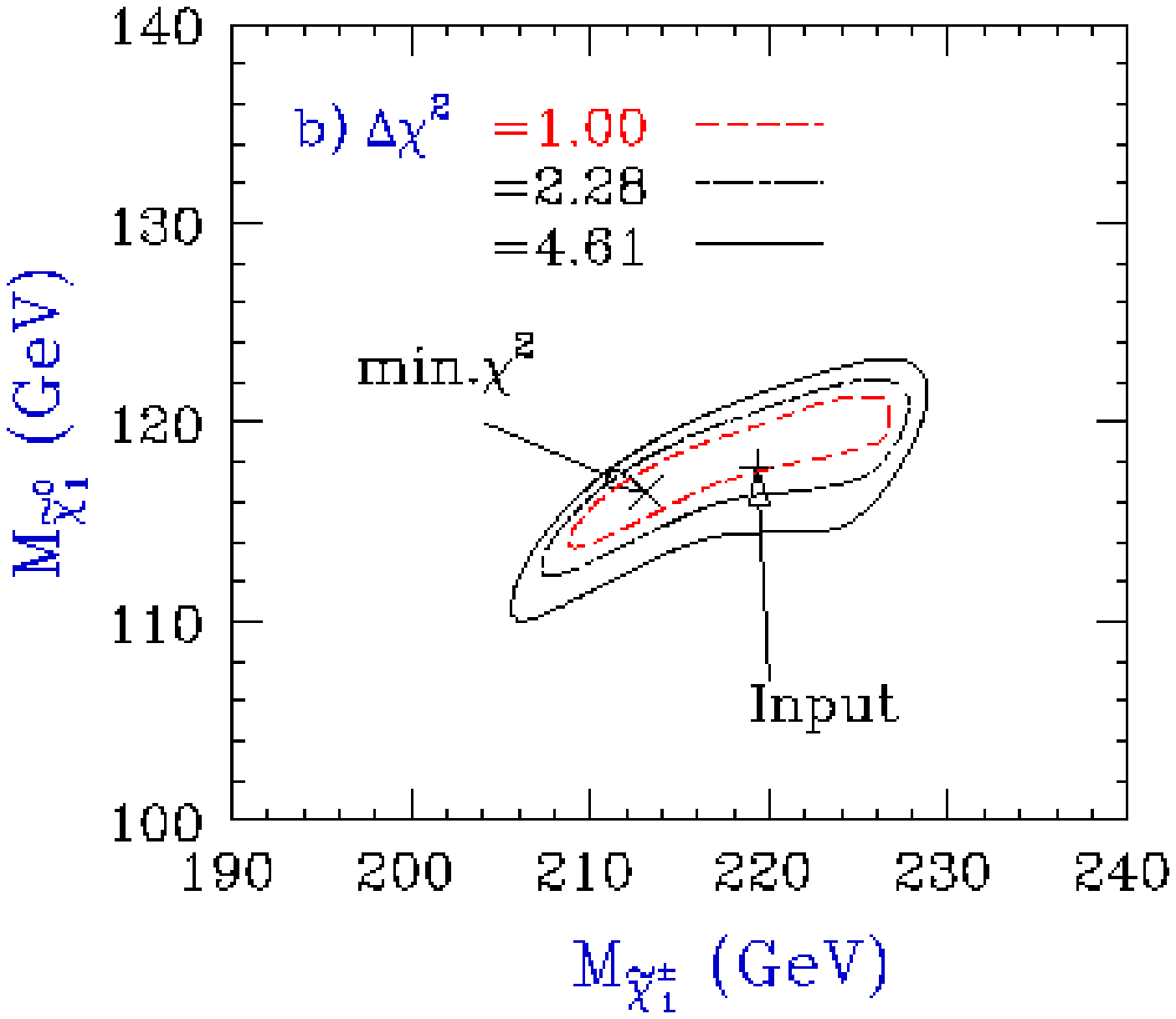}
 \includegraphics*[scale=0.35]{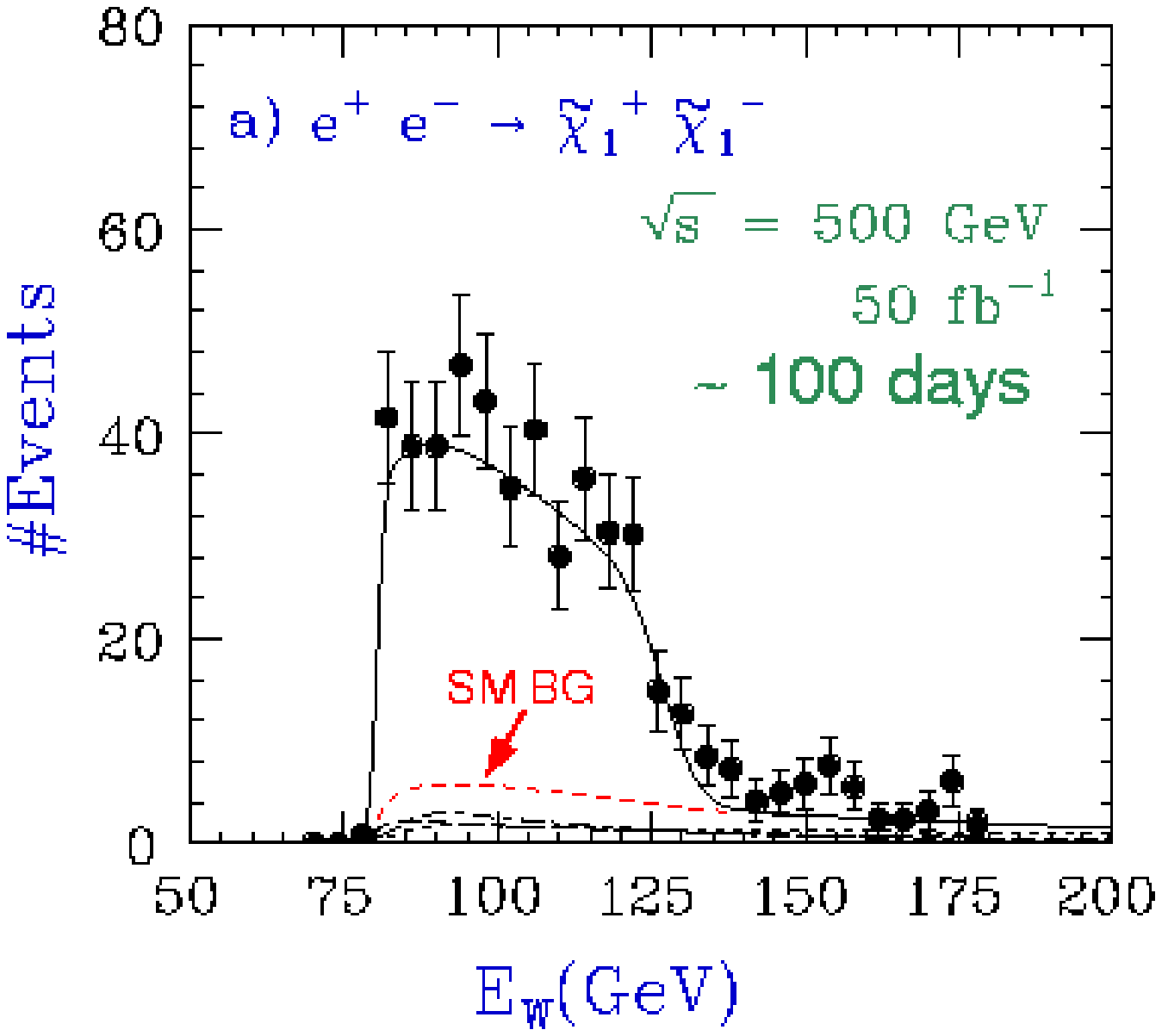}
}
\vspace {0.5in}
\centerline{
 \includegraphics*[scale=0.35]{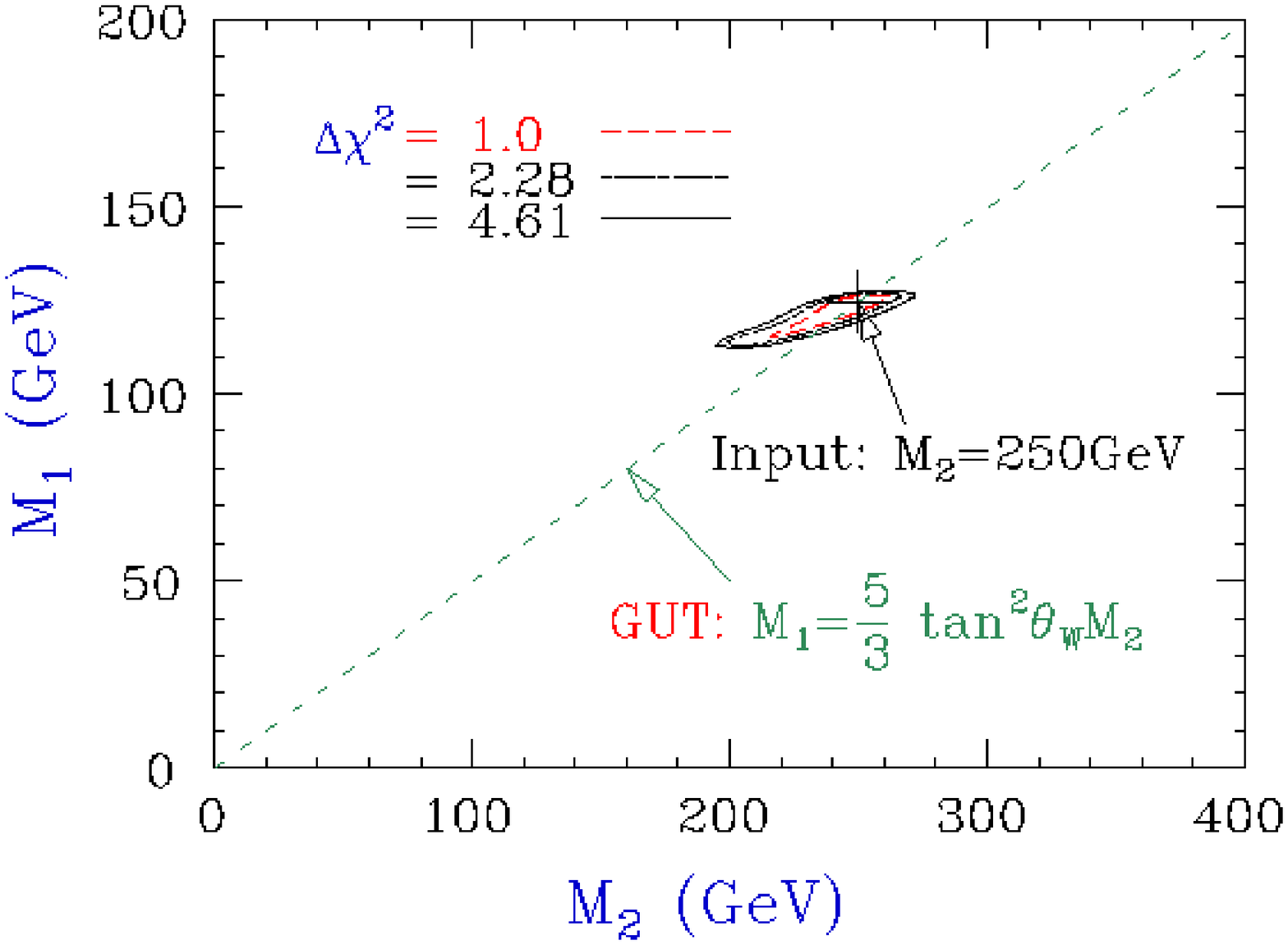}
}
\caption{Simultaneous determination of chargino/neutralino mass from 
chargino studies and consequent testing of the GUT relation between 
$M_1$ and $M_2$~\protect\cite{FM1}.}
\label{F:rgplen:13}
\end{figure}

Accurate high statistics measurements of the chargino system, provided 
both the charginos are accessible, also afford a good test of the equaility
of $g_{e \nu W}$ and $g_{e \snu \tilde W}$. For the representative
points in the SUGRA parameter space, chosen for the TESLA studies~\cite{Z3},
the relation can be tested to $0.1 \%$ for an integrated luminosity 
of $1~{\rm ab}^{-1}$. It should be noted however, that this study uses only 
the statistical errors in the analysis.
The above discussion thus shows that at an LC one can indeed measure the 
equality of the couplings which is the cleanest evidence for supersymmetry.

\section*{Determination of the SUSY breaking parameters at high scale and 
the SUSY breaking scale}
The precision measurements of the masses and the mixings in the sfermion and
the chargino/neutralino sector at the LC will certainly allow to establish
existence of supersymmetry as a dynamical symmetry of particle interactions.
However, this is not all these measurements can achieve. The high precision 
of these measurements will then allow us to infer about the SUSY breaking scale
and the values of the SUSY breaking parameters at this high scale, just the
same way the high precision measurements of the couplings $g_1,g_2$
and $g_3$ can be used to get a glimpse of the physics of unification and its
scale, as has been shown in Fig.~\ref{F:rgplen:0}.

There are essentially two different approaches to these studies. In the 
pioneering studies~~\cite{FM1,MIHO2}, the JLC group investigated how 
accurately one can determine the parameters $M_1,M_2,\mu,\tan\beta$ and $m_0$ 
{\it at the high scale} by fitting these {\bf directly} to the various 
experimental observables such as the polarisation dependent production
cross-sections of the sparticles, angular distributions of the decay products 
etc., that have been mentioned in the discussion so far. An example of this is
shown in Fig.~\ref{F:rgplen:13}. The right hand figure in the top panel 
shows how a determination of the energy distribution of the `W' produced in the
decay of $\Chip/\Chim $, in the reaction $\epem \to \Chip_1 \Chim_1$, affords a 
determination of $m_{\N0_1}$ and $m_{\Chipm}$ shown in the left panel. The
lower panel then shows how using the masses $m_{\Chipm},m_{\N0_1}$ along with
$\sigma_R(\Chimp_1 ), \sigma_R(\tilde e_R)$ and the angular distribution 
of the decay leptons one can extract $M_1,M_2$ at the GUT scale and test
the GUT relation.

\begin{figure}[htb]
\centerline{
 \includegraphics*[width=3.00in,height=2.00in]{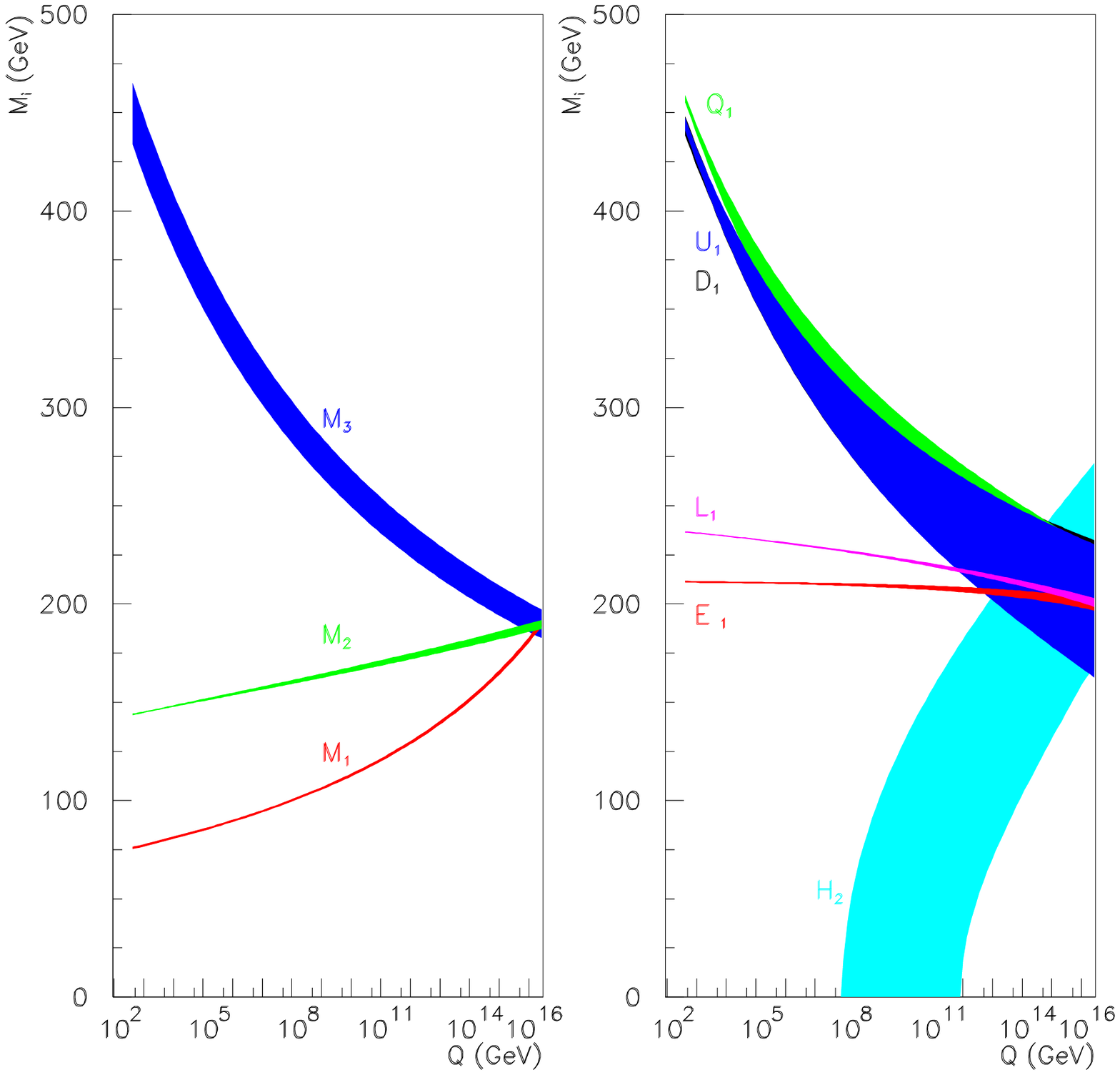}
 \includegraphics*[width=2.40in,height=2.00in]{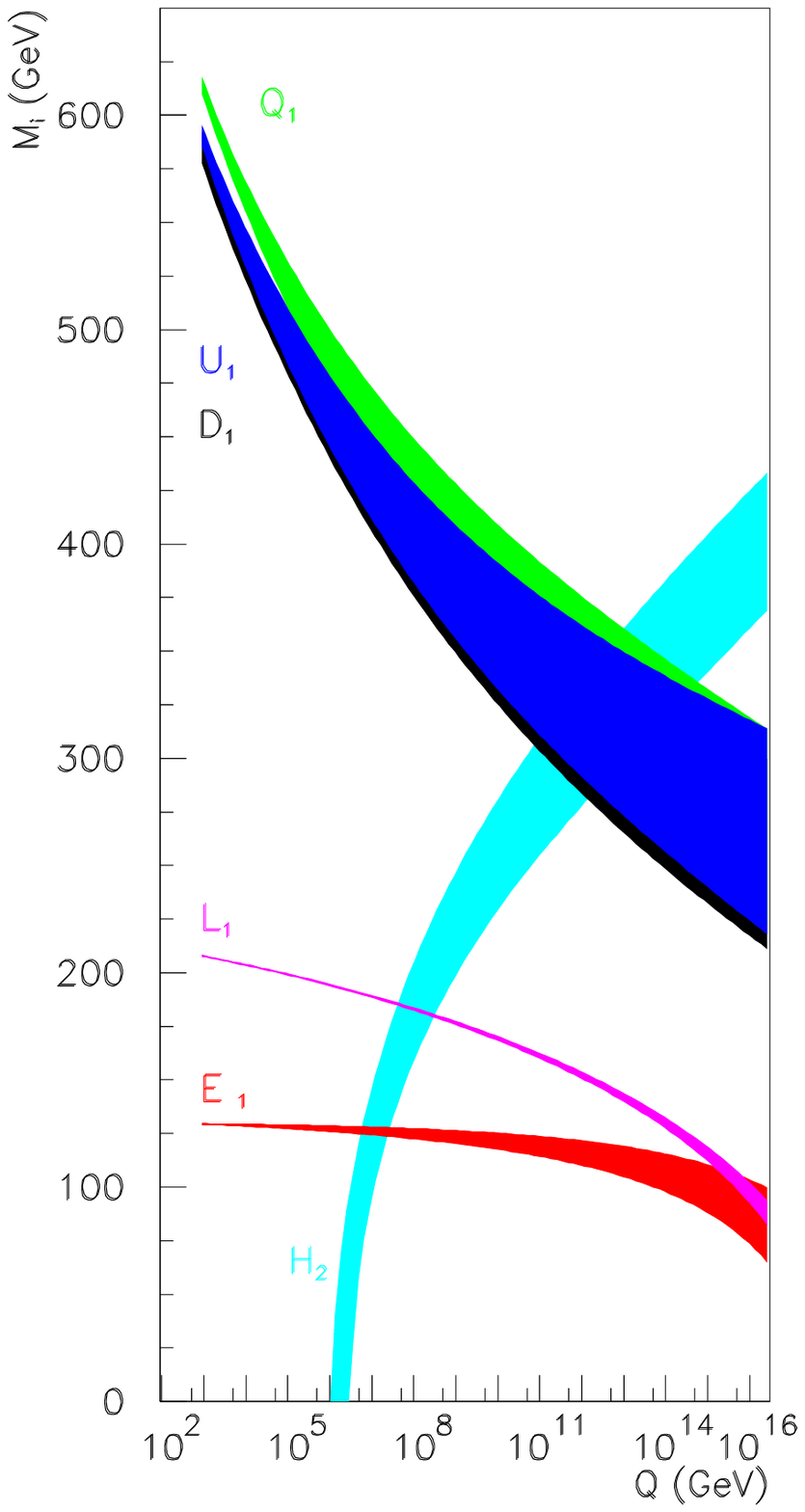}
}
\caption{Bottom up approach of the determination of the sparticle mass
parameters for mSUGRA and GMSB\protect\cite{PZB}. 
Values of the model parameters as given there.}
\label{F:rgplen:14}
\end{figure}

A different approach~\cite{zerwastlk,MB1,extra7,kneur1} is to use the 
experimental observables such as 
cross-sections, angular distributions to determine the physical parameters
of the system such as masses and mixings and then use these to determine the
Lagrangian parameters $M_1,M_2,\mu,\tan \beta$ at the {\it EW scale itself}. 
Thus the possible errors of measurements of the experimental quantities alone 
will  control the accuracy of the detrmination of these parameters. There are
again two ways in which this information can be used: one is a top down 
approach which in spirit is similar to the earlier one as now one uses these
accurately determined Lagrangian parameters at the EW scale to fit their values
at the high scale and then compare them with the input value. It has been 
shown~\cite{MB1} in this approach, that the projected accurate measurements 
of the 
various sparticle masses through threshhold scans with a very high luminosity 
run of TESLA (one will require a threshold scan using ten points with 
$10$\ ${\rm fb}^{-1}$ at each point, for each sparticle and appropriately 
higher energies for the heavier ones), allows a determination of the values of 
$M_1,M_2,m_0, \mu $ and $\tan\beta$ at the high scale to an accuracy of
better than $1 \% $. As mentioned earlier the accuracy is much worse for
higher values of $\tan \beta$. The expected accuracy of determination 
of the trilinear term is rather poor as shown in Tables~\ref{T:rgplen:3} and~\ref{T:rgplen:4} taken from Ref.~\cite{MB1}. This deterioration is due to 
the fact that most of 
the physical observables are rather insensitive to the parameters $A_k$.
\begin{table}
\noindent\begin{minipage}{0.5in}
$\mbox{ }$
\end{minipage}
\noindent\begin{minipage}{2in}
\caption{Reconstruction of SUGRA parameters assuming universal masses.}
\label{T:rgplen:3}
\centering
\begin{tabular}{ccc}
$\mbox{ }$ & True value & Error \\
\tableline
$m_0$ & 100 & 0.09 \\
$m_{1/2}$ & 200 & 0.10 \\
$A_0$ & 0 & 6.3 \\
tan$\beta$ & 3 & 0.02 \\
$\mbox{ }$ & $\mbox{ }$ & $\mbox{ }$ \\
\end{tabular}
\end{minipage}
\hfill
\noindent\begin{minipage}{2in}
\caption{Reconstruction of SUGRA parameters with nonuniversal guagino masses.}
\label{T:rgplen:4}
\centering
\begin{tabular}{ccc}
$\mbox{ }$ & True value & Error \\
\tableline
$m_0$ & 100 & 0.09 \\
$M_1$ & 200 & 0.20 \\
$M_2$ & 200 & 0.20 \\
$A_0$ & 0 & 10.3 \\
tan$\beta$ & 3 & 0.04 \\
\end{tabular}
\end{minipage}
\noindent\begin{minipage}{0.5in}
$\mbox{ }$
\end{minipage}
\end{table}
A completely different and a very interesting way of using the information on
these masses~\cite{PZB} is the bottom up approach where, one starts with these 
Lagrangian  parameters extracted at the weak scale and use the renormalisation 
group evolution (RGE) to calculate these parameters at the high scale. As 
explained in the introduction, different SUSY breaking mechanisms differ in 
their predictions for  relations among these various parameters at the high
scale. The ineteresting aspect of the bottom up approach is the possibility 
they offer of testing these relations `directly' by reconstructing them
from their low energy values using the RGE. In the analysis the
`experimental' values of the various sparticle masses are  generated in a
given scenario (mSUGRA,GMSB etc.) starting from the universal parameters
at the high scale appropriate for the model under consideration and using
the evolution from the high scale to the EW scale. These quantities are then
endowed with experimental errors expected to be reached in the combined 
analyses from LHC and an LC with energy upto $1$ TeV , with an integrated
luminosity of $1 {\rm ab}^{-1}$. Then these values are evloved once again
to the high scale. The  figure in the left panel of Fig.~\ref{F:rgplen:14}
shows  results of such an exercise for  the gaugino and sfermion masses for 
the mSUGRA case  and the one on the right for sfermion masses in GMSB. 
The width of the bands indicates $95 \%$ C.L.  Bear in mind 
that such accuracies will require a 10-20 year program at an LC 
with $\sqrt{s} \leq 1.5 $ TeV.

The two figures in the left panel show that with the projected 
accuracies of measurements, the unification of the gaugino masses will
be indeed demonstrated very clearly. The errors in the evolution of the slepton
masses are rather small as only the EW gauge couplings contribute to it. This
is seen in the second figure in left panel of the figure, which shows
the evolved squark, slepton and the Higgs masses. The unification of the slepton
masses at  the high scale can be demonstrated quite precisely. The 
errors on the reconstruction
of the universal mass from squark and higgs masses, are rather large. In the
case of the Higgs mass parameters this insenstivity to the common scalar mass 
$m_0$ is due to an accidental cancellation between different contributions
in the loop corrections to these masses which in turn control the RG evolution.
In the case of squarks the errors in the extrapolated values are caused by 
the stronger dependence of the radiative corrections on the common gaugino 
mass, due to the strong interactions of the squarks. As a result of these
a small error in the latter can magnify in the solution of $m_0$. Further, the 
trilinear $A$ coupling for the top shows a pseudo fixed point behaviour,
which again makes the  EW scale value insensitive to $m_0$. If the universal
gaugino mass $m_{1/2}$ is larger than the $m_0$ then this pesduo
fixed point behaviour increases the errors in the determination of 
third generation squark mass  at the EW scale.  This picture shows
us clearly the extent to which the unification at high scale can  be tested.
If we compare this with the results of Tables~\ref{T:rgplen:3} 
and \ref{T:rgplen:4}, we see that with the bottom up  approach we have a 
much clearer representation of the situation.
The $95 \%$ C.L. bands on the squark and the higgs mass parameters get much
wider if one assumes only the accuracies expected to be reached at 
the LHC collider~\cite{BWS}. Thus we see that an LC can help crucially 
in trying to  give a clearer picture of the SUSY breaking parameters at 
the high scale. 

The figure in the right  panel shows the results of a similar 
exercise  but for GMSB model, where with the assumed values of the model 
parameters, one would need to have a $1.5$ TeV LC to access the full 
sparticle spectrum. In this case the doublet slepton mass and the Higgs mass 
parameter is expected to unify at messenger scale which the `data' show 
quite clearly.  Further the high energy behaviour of the  reconstructed 
slepton and squark masses is accurate enough to see entirely different 
unification patterns expected in this case as opposed to the mSUGRA case. 
The bottom up approach of testing the strucutre of SUSY breaking parameters 
at high energy will work only with the high accuracy that one can reach 
at the LC. This point is discussed more at length (comparisons with
the results possible with the LHC measurements alone, other models
of SUSY breaking such as Gaugino mediated SUSY breaking etc.) elsewhere
in the proceedings~\cite{BWS}.

\begin{figure}[htb]
\centerline{
 \includegraphics*[scale=0.35]{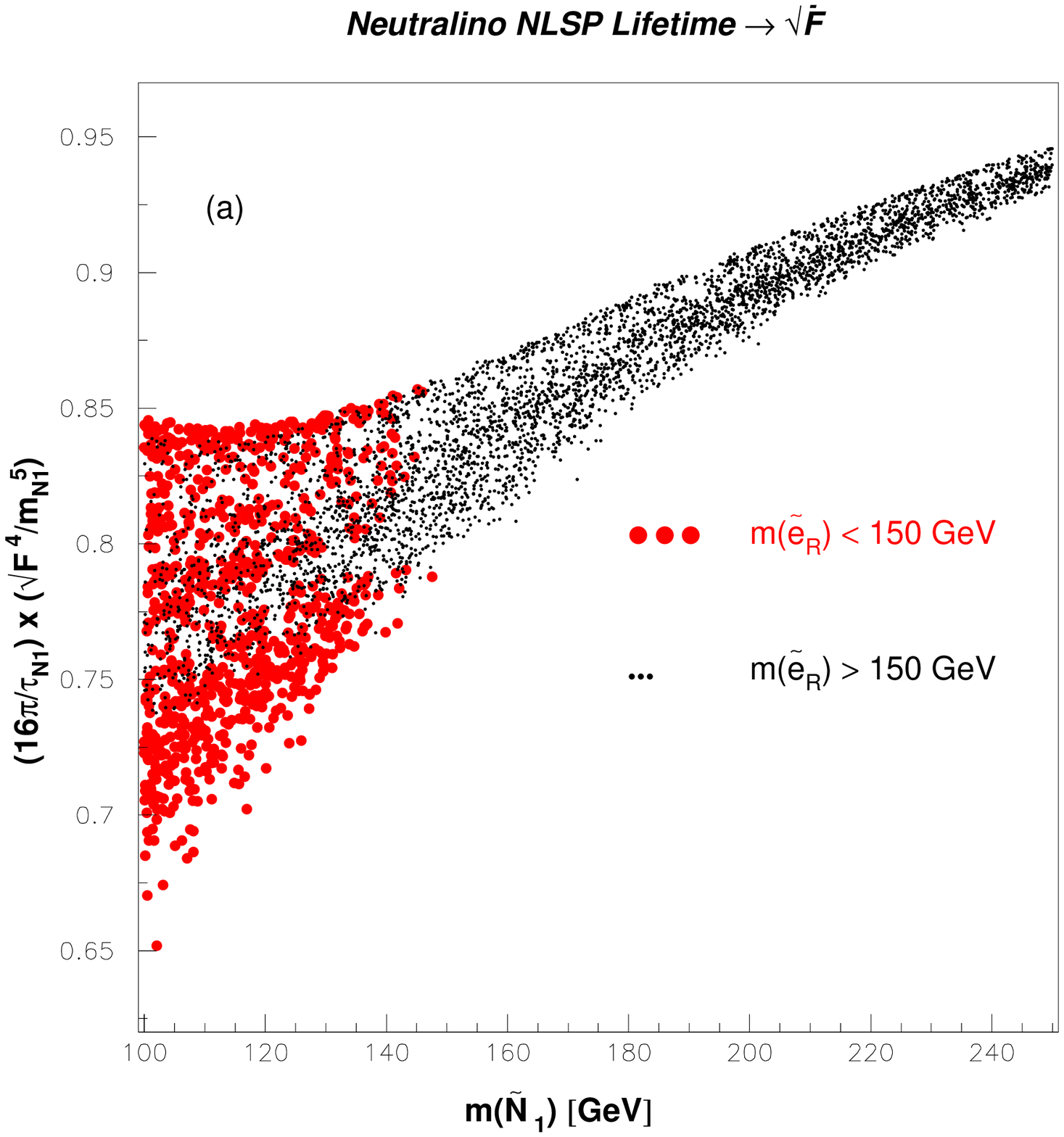}
 \includegraphics*[scale=0.35]{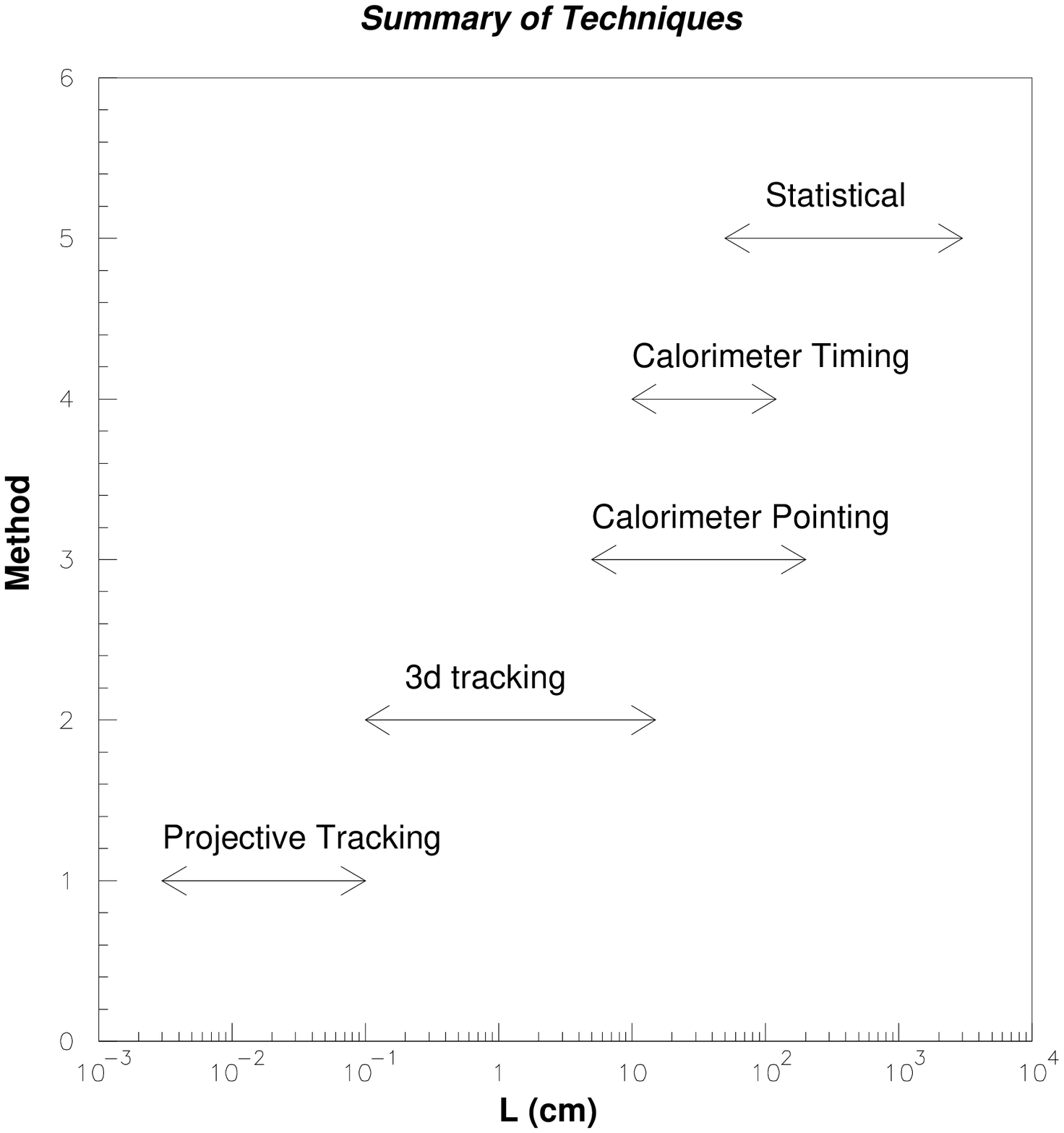}
}
\caption{Spread of theoretical expectations for Neutralino NLSP lifetime in 
GMSB models and a summary of techniques for its measurements.}
\label{F:rgplen:15}
\end{figure}

Since in the GMSB models the life time of the NLSP is determined
by the NLSP mass and the SUSY breaking scale, a measurement of the 
NLSP mass along with the decay life time can offer a very nice measurement
of the  breaking scale $\sqrt{F}$. The left panel  in Fig.~\ref{F:rgplen:15},
taken from Ref.~\cite{AB}, demonstrates that in the neutralino NLSP
scenario one has to a very good approximation
$$c \tau  \propto {1 \over {m_{\N0_1}^5}} (\sqrt{F})^4. $$
The figure shows the neutralino NLSP life time scaled with 
appropriate powers of the $m_{\N0_1}$ and $\sqrt{F}$ and one sees that it is 
a constant to within $10 \% $. The right panel in the same figure also 
shows various methods which can be employed to determine the
 decay length $L = c \tau$ with an accuracy of about $10 \% $. This 
shows that a determination of the NLSP decay length is possible in
the entire, rather large,  range $(10^{-3} - 10^{4}~{\rm cm})$ expected in 
the GMSB models.  The knowledge of the mass $m_{\N0_1}$ is crucial in 
this determination.  Thus this analysis demonstrates that if indeed GMSB
is realised in nature, the scale of breaking can be determined to 
within $10 \%$ , by a study of the NLSP neutralino at an LC, with a moderate
luminosity. Considering that the theoretical considerations allow it to lie 
in a rather wide range spanning three to four orders of magnitude, this 
would be a very interesting determination indeed.  The possibilities of
being able 
to tune the LC energy as well as the much cleaner environment available 
to measure the life time of the NLSP neutralino, allow a  much more accurate
measurement of $\sqrt{F}$ than is possible at  the LHC~\cite{AP,extra10}.

All these discussions assume that most of the sparticle spectrum
will be accessible jointly between the LHC and a TeV energy LC. 
If however, the squarks are superheavy~\cite{fengbagger,moroi} 
(in the focus point SUSY scenarios the entire scalar sector might be 
beyond  a few TeV), then perhaps the only
clue to their existence can be obtained through the analogue of precision
measurements of the oblique correction to the SM parameters at the Z pole. These
superoblique corrections~\cite{MIHO3}, modify the 
equalities between various couplings mentioned already in Eq.~\ref{eq:rg:1}.
These modifications arise if there is a large mass splitting between 
the sleptons and the squarks. The expected radiative corrections imply
\be
{{\delta g_Y} \over g_Y}  \simeq  {{11 g_Y^2} \over {48 \pi^2}} 
ln \left({m_{\sq}} \over {m_{\tilde l}}\right).
\ee 
Thus if the mass splitting is a factor 10 one expects a deviation from the
tree level relation by about $0.7 \% $. The discussions of the earlier
section demonstrate that it might be possible at an LC to make such a 
measurement. However, it must be mentioned that these statements are 
based on an anaylsis which essentially uses only  statistical errors. 
The effect of systematic errors on such measurements needs to be studied.
It has been shown, again using statistical errors alone, that at
an $e^-e^-$ collider,  with a study of the reaction 
$e^-e^- \to \sel^*_R \sel_R $,
one can determine these superoblique corrections to a much higher accuracy;
$\sim 0.15 \% $. This increase in the accuracy is possible because in this
reaction the $s$-channel diagram does not exist and only the $t$-channel 
diagram involving the $\N0_i$ exchange exists.

\section*{Conclusions}
The above discussion can be summarised very briefly by saying that a 
TeV scale, high luminosity  LC will not only be able to confirm the LHC
`discovery'  of TeV scale SUSY but will be able to  test coupling 
equalities expected in Supersymmtery thereby testing the most basic
prediction of the symmetry. It further will be able to provide precision 
infomation on SUSY, SUSY breaking scale and SUSY breaking mecahanism in a
`model independent' way. Such a collider will yield a lot of unambiguious
information about SUSY which is model independent and help discriminate SUSY 
in the signals for new physics we see, from  plausible alternative 
explanations  for  these that can be constructed~\cite{murWS} 
with enough ingenuity. 
The very interesting `bottom up' approach will require high
lumiosities  as high as $1~{\rm ab}^{-1}$ and energies upto
$1.5$ TeV, for reasonable particle spectra. At present, only the TESLA collider
designs~\cite{teslatdr}  envisage such high luminosities.
Most of the designs (apart from TESLA) are capable of extending upto the
$1.5$ TeV.  An LC with $500 < \sqrt{s} < 1500$ GeV should be capable of 
covering a major part of the range of model predictions for the sparticle 
masses. If some of these lie beyond the kinematical reach, measurements of the 
superoblique parameters should still be able to give us information on them.
The $e^-e^-$ colliders (which can be made reasonably easily, once
we have the $\epem$ machine) have special advantages when looking at the 
selectron pair production which, however, is the only SUSY channel 
available at these colliders.  As far as the $\gamma \gamma$ colliders
are concerned, the sparticle searches almost don't gain anything more over
what is possible at the corresponding $\epem$ collider, except the 
invstigations into the $H/A$ mixing in the $2$ Higgs doublet models such 
as SUSY. Just an LC running in the $\epem$ mode with $\sqrt{s} \leq 1500 $ GeV 
and $\int {\cal L} dt = 1\ {\rm ab}^{-1}$ will be sufficient to make
the precision measurements of SUSY as outlined above. Establishing 
the Lagrangian parameters of SUSY in such a manner will go a long way 
towards putting it in text books as `THE' theory of physics beyond the SM. 

\section* {Acknowledgements}
It is a pleasure to thank the organisers for an excellent meeting.

\end{document}